\documentclass[12pt]{iopart}
\expandafter\let\csname equation*\endcsname\relax
\expandafter\let\csname endequation*\endcsname\relax
\usepackage{graphicx,bm,citesort,amsmath,iopams,color}

\newcommand{\AJP}{ {\em Am. J. Phys. }}
\newcommand{\AnM}{ {\em Annals Math. }}
\newcommand{\APB}{ {\em Ann. Phys. (Berlin) }}
\newcommand{\CMP}{ {\em Commun. Math. Phys. }}
\newcommand{\CP}{ {\em Chem. Phys. }}
\newcommand{\CPl}{ {\em Chem. Phys. Lett. }}
\newcommand{\CRA}{ {\em C. R. Acad. Sci. Ser. A }}

\newcommand{\EPJP}{ {\em Eur. Phys. J. Plus }}

\newcommand{\IJQC}{ {\em Int. J. Quantum Chem. }}
\newcommand{\JMC}{ {\em J. Math. Chem. }}
\newcommand{\JSP}{ {\em J. Stat. Phys. }}

\newcommand{\LasP}{ {\em Laser Phys. }}

\newcommand{\PA}{ { \em Physica A }}
\newcommand{\PLA}{ {\em Phys. Lett. A }}
\newcommand{\PNAS}{ {\em P. Natl. Acad. Sci. USA }}
\newcommand{\PRA}{ {\em Phys. Rev. A }}
\newcommand{\PRB}{ {\em Phys. Rev. B }}

\newcommand{\PRD}{ {\em Phys. Rev. D }}

\newcommand{\PRSA}{ {\em Proc. R. Soc. A }}

\begin{document}
\title[One-dimensional pseudoharmonic oscillator]{One-dimensional pseudoharmonic oscillator: classical remarks and quantum-information theory}
\author{O Olendski}\address{Department of Applied Physics and Astronomy, University of Sharjah, P.O. Box 27272, Sharjah, United Arab Emirates}
\ead{oolendski@sharjah.ac.ae}

\begin{abstract}
Motion along semi-infinite straight line in a potential that is a combination of positive quadratic and inverse quadratic functions of the position is considered with the emphasis on the analysis of its quantum-information properties. Classical measure of symmetry of the potential is proposed and its dependence on the particle energy and the factor $\mathfrak{a}$ describing a relative strength of its constituents is described; in particular, it is shown that a  variation of the parameter $\mathfrak{a}$ alters the shape from the half-harmonic oscillator (HHO) at $\mathfrak{a}=0$ to the perfectly symmetric one of the double frequency oscillator (DFO) in the limit of huge $\mathfrak{a}$. Quantum consideration focuses on the analysis of information-theoretical measures, such as standard deviations, Shannon, R\'{e}nyi and Tsallis entropies together with Fisher information, Onicescu energy and non--Gaussianity. For doing this, among others, a method of calculating momentum waveforms is proposed that results in their analytic expressions in form of the confluent hypergeometric functions. Increasing parameter $\mathfrak{a}$ modifies the measures in such a way that they gradually transform into those corresponding to the DFO what, in particular, means that the lowest orbital saturates Heisenberg, Shannon, R\'{e}nyi and Tsallis uncertainty relations with the corresponding position and momentum non--Gaussianities turning to zero. A simple expression is derived of the orbital-independent lower threshold of the semi-infinite range of the dimensionless R\'{e}nyi/Tsallis coefficient where momentum components of these one-parameter entropies exist which shows that it varies between $1/4$ at HHO and zero when $\mathfrak{a}$ tends to infinity. Physical interpretation of obtained mathematical results is provided.
\end{abstract}
\vskip.7in

\noindent

\submitto{Journal of Physics Communications}
\maketitle

\section{Introduction}\label{Sec_Intro}
It is well known \cite{Davidson1,Sage1,Oyewumi1,Yahya1} that a three-dimensional (3D) spherically symmetric potential
\begin{equation}\label{Potential3D_1}
V^{(3D)}(r)=D_\omega\left(\frac{r}{r_\omega}-\frac{r_\omega}{r}\right)^2
\end{equation}
written in position coordinates ${\bf r}=(r,\theta_{\bf r},\varphi_{\bf r})$ describes quite well roto-vibrational phenomena in diatomic molecules, with $D_\omega$ and $r_\omega$ being the dissociation energy between two atoms and equilibrium bond length, respectively. Its slightly amended 2D version
\begin{equation}\label{Potential2D_1}
V^{(2D)}(\mathfrak{a};r)=\frac{1}{2}m\omega^2r^2+\frac{\hbar^2}{2mr^2}\,\mathfrak{a}-\hbar\omega\mathfrak{a}^{1/2}=\frac{1}{4}\hbar\omega\left(\frac{r}{r_\omega}-2\mathfrak{a}^{1/2}\frac{r_\omega}{r}\right)^2,
\end{equation}
is indispensable in the analysis of the flat quantum rings \cite{Bogachek1,Tan1,Tan2,Tan3,Fukuyama1,Bulaev1,Simonin1,Olendski5,Gumber1,Olendski1,Olendski2}. Here, $m$ is a mass of the carrier, frequency $\omega$ defines a steepness of the confining in-plane outer surface of the annulus with its characteristic radius $r_\omega=[\hbar/(2m\omega)]^{1/2}$, and positive dimensionless constant $\mathfrak{a}$ describes a strength of the repulsive potential near the origin. 1D analogue of the above expressions for the straight motion along the positive $x$ semi-axis
\begin{equation}\label{Potential1D_1}
V^{(1D)}(\mathfrak{a};x)=D_\omega\left(\frac{x}{x_\omega}-\mathfrak{a}^{1/2}\frac{x_\omega}{x}\right)^2,\quad0\leq x<\infty,
\end{equation}
was first introduced (with the simplifying assumption $\mathfrak{a}=1$) more than sixty years ago in a problem-solving textbook on quantum mechanics \cite{Goldman1}. Despite of a such venerable age \cite{Weissman1,Nieto1,Gutschick1,Weissman2,Nieto2,Sage2,Ballhausen1,Ballhausen2,Ballhausen3,Hall1,Crawford1,Hall2,Chalykh1,Dong1,Dong2,Tavassoly1,Mikulski1,Baykal1}, some  properties of the movement in the similar type potentials that are commonly referred to as pseudoharmonic oscillator (PHO) still have not been addressed; namely, whereas quantum information measures, such as Shannon \cite{Shannon1,Shannon2}, R\'{e}nyi \cite{Renyi1,Renyi2} and Tsallis \cite{Tsallis1} entropies, Fisher information \cite{Fisher1,Frieden1}, Onicescu energy \cite{Onicescu1}, have recently been calculated for the 2D \cite{Olendski1,Olendski2} and 3D \cite{Yahya1} PHO, their 1D counterparts still almost have not been tackled with: the only exception is the discussion of the position and momentum Shannon and position Fisher functionals for the half harmonic oscillator (HHO) \cite{Shi1}, i.e., the structure with $\mathfrak{a}=0$. The primary aim of the present exposition is to close this gap. Mentioned above measures play a very important role in quantum information theory and it is essential for everyone dealing with this subject to have (at least superficial) knowledge about them. Experimental advances in evaluating, e.g., the Shannon \cite{Lukin1,Niknam1} and R\'{e}nyi \cite{Niknam1,Islam1,Kaufman1,Brydges1} entropies force the researchers to better understand their theoretical foundations what will facilitate an implementation of these quantities in quantum-information processing. From this point of view, a consideration of the relatively simple models that allow manageable {\it analytic} expressions for the above-mentioned functionals sheds new light on their properties and, accordingly, enriches our knowledge about them. The main focus of our research aims at the influence of the variation of the dimensionless coefficient $\mathfrak{a}$ on the classical and quantum properties of the structure. As shown below, the 1D geometry with the potential from equation~\eref{Potential1D_1} provides {\it analytic} dependencies of standard deviations on $\mathfrak{a}$ for any orbital in either position or momentum space whereas for the Fisher information it is done for the position  component only with all other measures exhibiting evident  formulas just for the ground-state position parts. A cornerstone result on which the whole subsequent analysis is based highlights that an alteration of the parameter transforms the shape described by equation~\eref{Potential1D_1} from the highly asymmetric potential at $\mathfrak{a}=0$ where the minimum of the HO with frequency $\omega=\left(2D_\omega/m\right)^{1/2}/x_\omega$ serves simultaneously as its left infinitely high boundary to the symmetric with respect to its only extremum regular HO with the frequency $2\omega$ -- in the opposite limit of the very large $\mathfrak{a}$. In this process, the lowest-energy waveforms in either space undergo transformation from non--Gaussian to Gaussian dependencies. Non--Gaussianity is an important property of the quantum systems that finds its applications, among others, in quantum-information protocols, such as, e.g., entanglement distillation and entanglement swapping \cite{Walschaers1}. To quantify it, below the measure based on the relative Shannon entropy \cite{Genoni1,Genoni2,Paris1,Albarelli1} is used and its components' decaying at the huge $\mathfrak{a}$ is analyzed. Simultaneously, the non--Gaussianity based on the quantum von Neumann entropy is calculated and compared to its position and momentum counterparts. Since, at the same time, the HO ground level at $\mathfrak{a}\rightarrow\infty$ saturates Heisenberg, Shannon, R\'{e}nyi and Tsallis uncertainty relations, from fundamental point of view it is extremely important and instructive to track and describe the approach of the corresponding measures to their HO counterparts when, e.g.,  the product of the position and wave vector deviations or the sum of the corresponding Shannon parts come closer and closer to the basal limits of one half and $1+\ln\pi$, respectively. This is why below the main attention is devoted (but not limited) to the lowest-energy orbital.

Simultaneously with the vigorous research in the field, the leading Universities actively include quantum-information theory into their curricula \cite{Aiello1,Cervantes1} and for other academical institutions it seems in the nearest future an inevitable but welcome reality \cite{Fox1} what forces the educators to seek didactic and methodological ways to discuss with the students physical and mathematical basics of the corresponding functionals. Lately, several such attempts aimed at the graduate and senior undergraduate level have been made \cite{Saha1,Olendski3,Olendski4,Nascimento1}. Postgraduate courses on a more general topic of 'the concept of information in physics' are being offered not only to the physicists but to a wider audience of "students of other natural sciences as well as mathematics, informatics, engineering, sociology, and philosophy" \cite{Dittrich1}. In this context, the present exposition might be considered as a continuation of those previous pedagogical efforts.

The outline of presentation is as follows. Before addressing quantum aspects, Sec.~\ref{Sec_Classical} that considers a classical motion shows that some quantities such as, e.g., periodic time, distance between the two turning points and average speed, stay unaffected by $\mathfrak{a}$. The first of them is not influenced by the particle energy $E$ either what is a consequence of the fact that the two latter ones obey a square root  dependence on $E$. A classical measure of symmetry is proposed that is the ratio of the distances between the zero potential position and the left and right turning points, respectively. This quantity monotonically increases from zero for HHO to its asymptotic value of unity at the very large $\mathfrak{a}$. Initial step of Sec.~\ref{Sec_Quantum} describes quantized energy spectrum and the corresponding position waveforms, which are the eigen values and  eigen vectors of the time-independent Schr\"{o}dinger equation. Analytic method of calculating momentum functions is proposed and its advantages are discussed. All this knowledge serves as a foundation to the main part that copes with the miscellaneous functionals of the position and momentum densities and their $\mathfrak{a}$ dependence. Measures addressed are: standard deviations, Shannon, R\'{e}nyi and Tsallis entropies, Fisher informations, Onicescu energies and non--Gaussianities.  Since their physical significance is described in many sources, including pedagogical ones \cite{Saha1,Olendski3,Olendski4,Nascimento1,Griffiths1}, they are only briefly considered. For the first four of them, the uncertainty relations between the position and wave vector components are known and, as shown below, for the ground level they saturate with the growth of $\mathfrak{a}$ since the squares of the corresponding waveforms come closer and closer to the Gaussian dependencies. Discussion of the momentum wave function that started in section~\ref{SubSec_Energy} and revealed that its expression contains a finite sum of confluent hypergeometric functions is employed for a derivation of a simple formula for the lowest limit of the semi-infinite range of the R\'{e}nyi and Tsallis coefficients where these momentum functionals make sense. Derived mathematical formulas received their physical interpretation; in particular, a special role of the HO is highlighted.

\begin{figure}
\centering
\includegraphics[width=0.6\columnwidth]{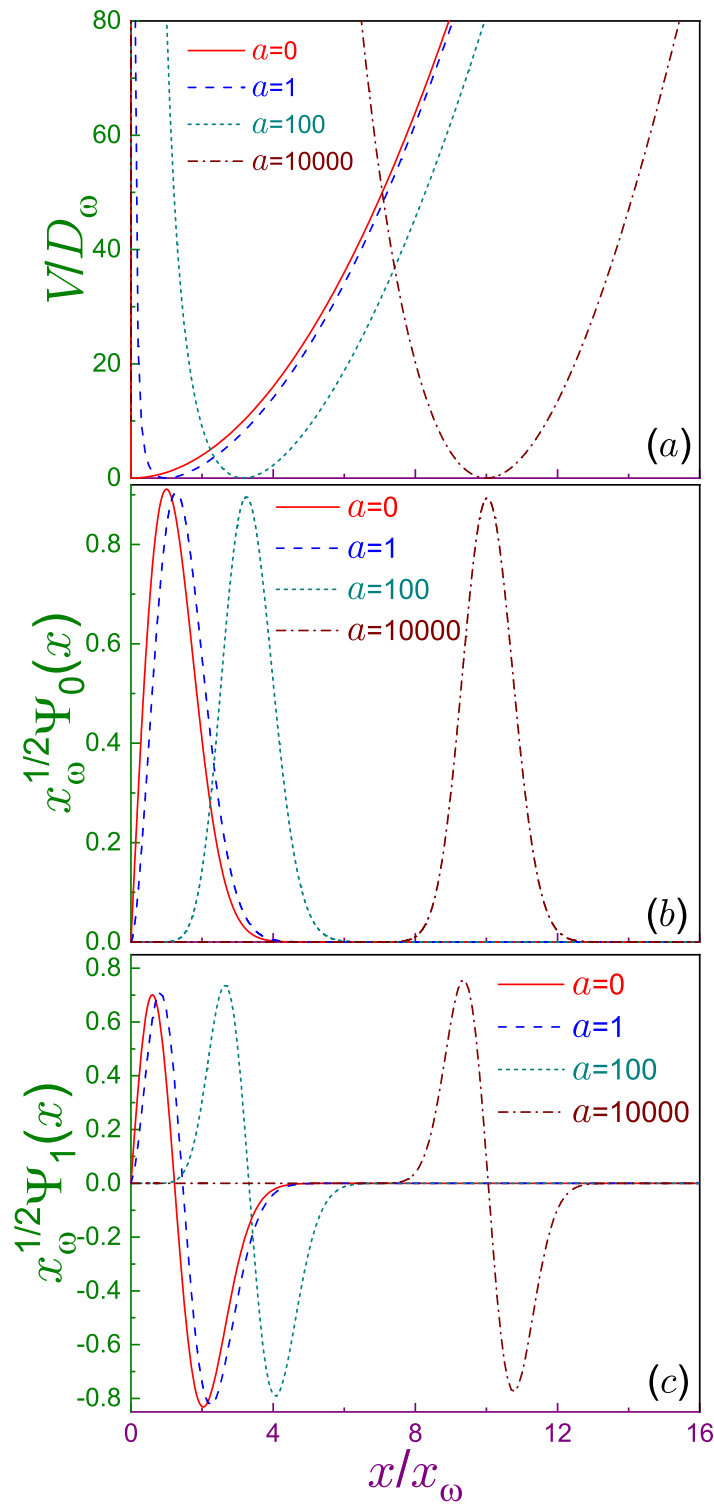}
\caption{\label{Fig_PotentialPositionWaveforms}
(a) Dimensionless potential profile from equation~\eref{Potential1D_1} at several parameters $\mathfrak{a}$ in terms of the normalized position $x/x_\omega$. (b) Ground and (c) first-excited position wave functions (scaled by $x_\omega^{-1/2}$) at the same $\mathfrak{a}$ as those in window (a). For all subplots, solid lines correspond to $\mathfrak{a}=0$, dashed ones are for $\mathfrak{a}=1$, dotted curves depict the case of $\mathfrak{a}=100$ and dash-dotted dependencies describe the geometry with $\mathfrak{a}=10000$. Note different vertical ranges in panels (b) and (c).}
\end{figure}

\section{Potential profile and classical motion}\label{Sec_Classical}
Throughout the whole discussion, we will interchangeably use the energy $D_\omega=\frac{1}{2}m\omega^2x_\omega^2$ and its quantum equivalent $\hbar\omega/2$ where the confining frequency, which in classical mechanics is determined by the force constant $\cal K$ as $\omega=\sqrt{{\cal K}/m}$, defines the length $x_\omega$ according to
\begin{equation}\label{Xomega}
x_\omega=\sqrt{\frac{\hbar}{m\omega}}.
\end{equation}

Figure~\ref{Fig_PotentialPositionWaveforms}(a) exhibits the shape of the potential profile, equation~\eref{Potential1D_1}, for several parameters $\mathfrak{a}$. Its zero minimum is achieved at 
\begin{equation}\label{PositionZero1}
x_Z=\mathfrak{a}^{1/4}x_\omega.
\end{equation}
Around this only extremum, at the positive coefficient, $\mathfrak{a}>0$, its behaviour is governed by the Taylor expansion:
\begin{eqnarray}
V^{(1D)}(\mathfrak{a};x)&=4D_\omega\left(\frac{x}{x_\omega}-\mathfrak{a}^{1/4}\right)^2\left[1-\frac{1}{\mathfrak{a}^{1/4}}\left(\frac{x}{x_\omega}-\mathfrak{a}^{1/4}\right)\right.\nonumber\\
\label{PotentialLimit1}
&\left.+\frac{5}{4\mathfrak{a}^{1/2}}\left(\frac{x}{x_\omega}-\mathfrak{a}^{1/4}\right)^2-\ldots\right],\quad\frac{1}{\mathfrak{a}^{1/4}}\left|\frac{x}{x_\omega}-\mathfrak{a}^{1/4}\right|<1.
\end{eqnarray}
At $\mathfrak{a}=0$, equation~\eref{Potential1D_1} describes the geometry of a particle confined to the right-hand half of the HO of the frequency $\omega$. Increasing repulsive factor $\mathfrak{a}$ shifts the minimum of the potential to the right, as it follows from equation~\eref{PositionZero1}, simultaneously transforming it at the greater $\mathfrak{a}$ into the more symmetric shape: the leading term of equation~\eref{PotentialLimit1} corresponds to the HO with frequency $2\omega$ centered at $x_Z$ with the influence of the subsequent items there decreasing in the limit $\mathfrak{a}\rightarrow\infty$.

\begin{figure}
\centering
\includegraphics[width=\columnwidth]{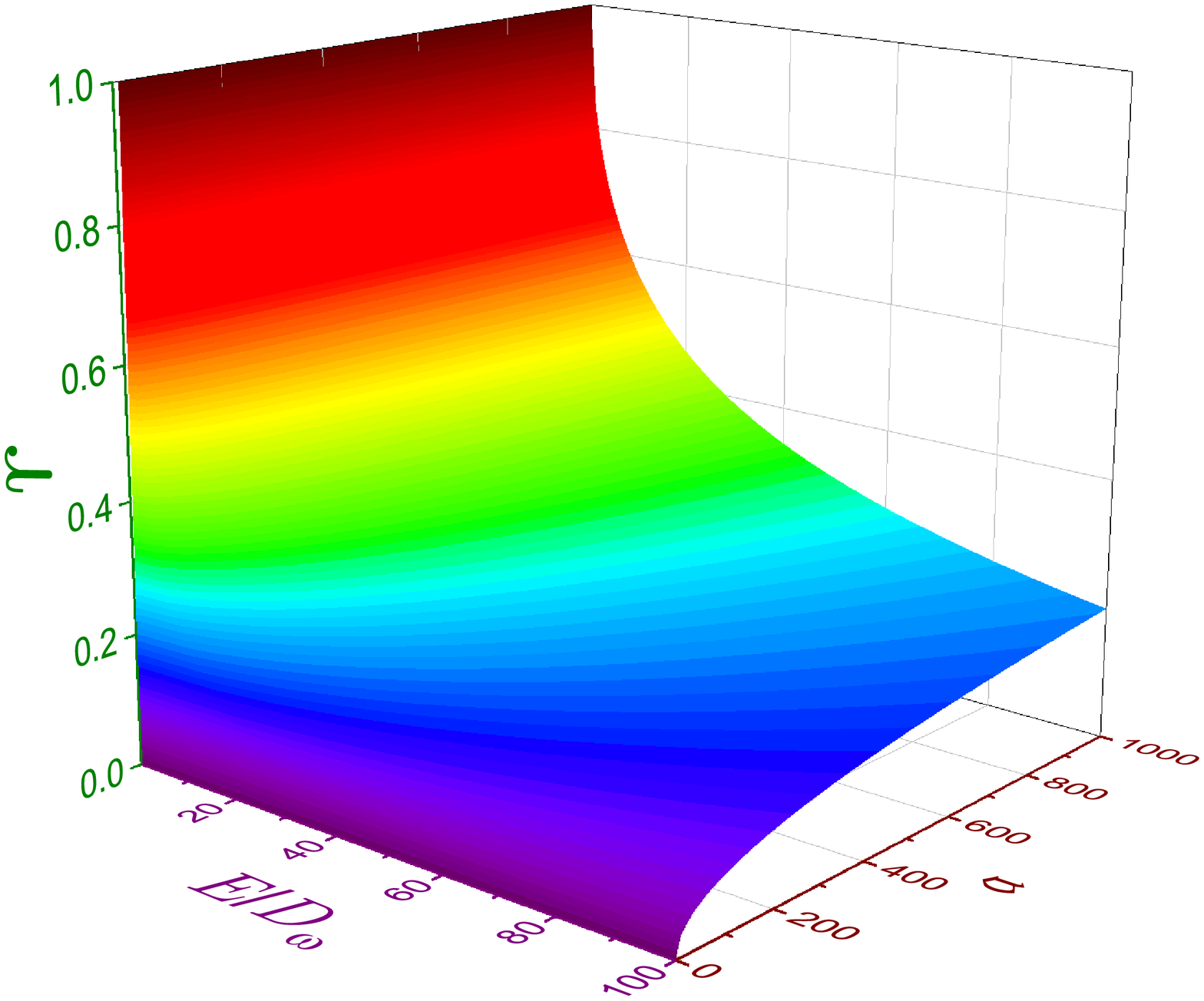}
\caption{\label{Fig_ClassicalSymmetry}
Classical symmetry factor $r$, equation~\eref{Asymmetry1}, in terms of parameter $\mathfrak{a}$ and ratio $\frac{E}{D_\omega}$.}
\end{figure}

Symmetrization of the potential with the growing $\mathfrak{a}$ is confirmed from the classical analysis; namely, for the particle with the energy $E$, its motion is confined between the left $x_-$ and right $x_+$ turning points:
\begin{equation}\label{TurningPoints1}
x_\pm=x_\omega\sqrt{\frac{E}{2D_\omega}+\mathfrak{a}^{1/2}\pm\sqrt{\left(\frac{E}{2D_\omega}\right)^2+\frac{E}{D_\omega}\,\mathfrak{a}^{1/2}}}
\end{equation}
with the asymptotes:
\begin{subequations}\label{TurningPoints2}
\begin{align}\label{TurningPoints2_SmallA}
&\left.\begin{array}{l}
x_-=\left(\frac{D_\omega}{E}\right)^{1/2}\mathfrak{a}^{1/2}\left[1-\frac{D_\omega}{E}\mathfrak{a}^{1/2}+2\left(\frac{D_\omega}{E}\right)^2\mathfrak{a}-\ldots\right]x_\omega\\
x_+=\left(\frac{E}{D_\omega}\right)^{1/2}\left[1+\frac{D_\omega}{E}\mathfrak{a}^{1/2}-\left(\frac{D_\omega}{E}\right)^2\mathfrak{a}+\ldots\right]x_\omega
\end{array} \right\},\quad\frac{E}{D_\omega}\gtrsim1,\quad\mathfrak{a}\rightarrow0\\
\label{TurningPoints2_LargeA}
&x_\pm=\left[\mathfrak{a}^{1/4}\pm\frac{1}{2}\left(\frac{E}{D_\omega}\right)^{1/2}+\frac{1}{8\mathfrak{a}^{1/4}}\frac{E}{D_\omega}-\ldots\right]x_\omega,\quad\frac{E}{D_\omega}\lesssim1,\quad\mathfrak{a}\rightarrow\infty.
\end{align}
\end{subequations}
As a measure of symmetry, one can use the ratio
\begin{equation}\label{Asymmetry1}
r\!\left(\mathfrak{a},\frac{E}{D_\omega}\right)=\frac{\Delta_-}{\Delta_+}
\end{equation}
with $\Delta_\pm$ being the distances between the position of the minimum and right/left turning points:
\begin{equation}\label{DeltaPM}
\Delta_\pm=|x_\pm-x_Z|.
\end{equation}
Its limits are:
\begin{equation}\label{Asymmetry2}
r=\left\{\begin{array}{cl}
\mathfrak{a}^{1/4}\left(\frac{D_\omega}{E}\right)^{1/2}\left[1-\frac{D_\omega}{E}\mathfrak{a}^{1/2}+2\left(\frac{D_\omega}{E}\right)^2\mathfrak{a}-\ldots\right],&\frac{E}{D_\omega}\gtrsim1,\quad\mathfrak{a}\rightarrow0\\
1-\frac{1}{2\mathfrak{a}^{1/4}}\left(\frac{E}{D_\omega}\right)^{1/2}+\frac{1}{8\mathfrak{a}^{1/2}}\frac{E}{D_\omega}-\ldots,&\frac{E}{D_\omega}\lesssim1,\quad\mathfrak{a}\rightarrow\infty
\end{array} \right..
\end{equation}
Vanishing ratio at $\mathfrak{a}$ tending to zero is naturally explained by the potential minimum in this regime moving closer and closer to the left turning point and actually merging with it for HHO, $\mathfrak{a}=0$, whereas in the opposite limit of the huge parameter the zero-velocity positions are located almost symmetrically with respect to $x_Z$. On the other hand, since at the zero energy the particle does not move, both $x_+$ and $x_-$ are equal to $x_Z$ what mathematically directly follows from equations~\eref{TurningPoints1} and~\eref{PositionZero1}, and, accordingly, the ratio of the two zero values of $\Delta_-$ and $\Delta_+$ yields the unity symmetry factor, $r(\mathfrak{a},0)=1$. Dependence $r\!\left(\mathfrak{a},\frac{E}{D_\omega}\right)$ is depicted in figure~\ref{Fig_ClassicalSymmetry}. Observe that at any $\mathfrak{a}$ the symmetry  decreases with the energy growing.

Closely related to the above topic is a concept of a periodic time $T$ which is a duration of a travel from, e.g., the left turning point towards its right counterpart, bouncing at it and a subsequent return to the initial position $x_-$ \cite{Landau1}:
\begin{equation}\label{Time1}
T(E)=\sqrt{2m}\int_{x_-}^{x_+}\frac{dx}{\sqrt{E-V^{(1D)}(\mathfrak{a};x)}},
\end{equation}
where the argument at $T$ underlines that in general this quantity is a function of energy. Plugging in the 1D potential from equation~\eref{Potential1D_1} into this formula and calculating the integral with its limits from equation~\eref{TurningPoints1} yields:
\begin{subequations}\label{DiameterTimePHO1}
\begin{align}\label{Time2}
T^{PHO}&=\frac{\pi}{\omega}.
\intertext{This result deserves some attention. First, it manifests that for our shape the periodic time does \textit{not} depend on $E$. Potentials obeying such condition are called isochronous \cite{Calogero1}. It is known that to be isochronous, the potential needs to guarantee that its diameter $x_+-x_-$ should vary with energy as its square root \cite{Pippard1}. Calculating $(x_+-x_-)^2$ with the help of equation~\eref{TurningPoints1}, one elementary obtains that it is really the case for the PHO:}
\label{Diameter1}
x_+-x_-&=\left(\frac{E}{D_\omega}\right)^{1/2}\!\!\!x_\omega=\frac{1}{\omega}\left(\frac{2E}{m}\right)^{1/2}.
\end{align}
\end{subequations}
It was shown that a combination of the quadratic and inverse quadratic dependencies like the one from equation~\eref{Potential1D_1} (up to a shift and adding a constant) is the only 1D rational potential satisfying isochronous requirement \cite{Chalykh1}. Second, the diameter $x_+-x_-$, equation~\eref{Diameter1}, and, as a consequence, the periodic time $T$, equation~\eqref{Time2}, of our system are not influenced by the parameter $\mathfrak{a}$ either and, moreover, they are equal to the one half of their HO counterparts with frequency $\omega$:
\begin{subequations}\label{DiameterTimeHO1}
\begin{align}\label{DiameterTimeHO1_Time}
T_\omega^{HO}&=\frac{2\pi}{\omega}\\
\label{DiameterTimeHO1_Diameter}
x_+^{HO}-x_-^{HO}&=\frac{2}{\omega}\left(\frac{2E}{m}\right)^{1/2};
\end{align}
\end{subequations}
in other words, regardless of the coefficient $\mathfrak{a}$, the periodic time of the potential, equation~\eref{Potential1D_1}, is equal to that of the HO with frequency $2\omega$. It is the easiest to explain these results in the limiting cases of the strong and weak $\mathfrak{a}$; e.g., for the HHO, $\mathfrak{a}=0$, the distance traveled along one closed loop is precisely two times shorter than for the HO with the shape of the cut part being a mirror symmetric of that at $x>0$ what results in halving the time. Note that the average speed $s_{avg}$ for both potentials is the same:
\begin{equation}\label{AverageSpeed1}
s_{avg}^{PHO}=s_{avg}^{HO}=\frac{2}{\pi}\left(\frac{2E}{m}\right)^{1/2}.
\end{equation}

\section{Quantum-information measures}\label{Sec_Quantum}
\subsection{Energy spectrum and  position and momentum wave functions}\label{SubSec_Energy}
A first step in the quantum analysis is a finding of the energies $E_n$ and position waveforms $\Psi_n(x)$, $n=0,1,\ldots$, which are eigen values and eigen functions of the 1D Schr\"{o}dinger equation:
\begin{equation}\label{Schrodinger1}
-\frac{\hbar^2}{2m}\frac{d^2}{dx^2}\Psi_n(x)+V^{(1D)}(x)\Psi_n(x)=E_n\Psi_n(x).
\end{equation}
For the $\mathfrak{a}$-dependent potential from equation~\eref{Potential1D_1}, they are:
\begin{subequations}\label{ShcrodingerSol1}
\begin{align}\label{ShcrodingerSol1_Energy}
E_n(\mathfrak{a})&=\hbar\omega\left(2n+1+\eta-\mathfrak{a}^{1/2}\right)\\
\label{ShcrodingerSol1_Psi}
\Psi_n(\mathfrak{a};x)&=\frac{1}{x_\omega^{1/2}}\left[\frac{2n!}{\Gamma(n+\eta+1)}\right]^\frac{1}{2}\left(\frac{x}{x_\omega}\right)^{\eta+\frac{1}{2}}\exp\left(-\frac{1}{2}\frac{x^2}{x_\omega^2}\right)L_n^{(\eta)}\left(\frac{x^2}{x_\omega^2}\right).
\end{align}
\end{subequations}
Here, $\eta=\frac{1}{2}\sqrt{1+4\mathfrak{a}}$, $\Gamma(z)$ is $\Gamma$-function \cite{Abramowitz1}, $L_n^{(\alpha)}(z)$ is $n$th order associated Laguerre polynomial \cite{Abramowitz1}, and the set $\Psi_n(\mathfrak{a};x)$ at any parameter $\mathfrak{a}$ is orthonormalized:
\begin{equation}\label{OrthoNormalization_Position1}
\int_0^\infty\Psi_n(\mathfrak{a};x)\Psi_{n'}(\mathfrak{a};x)dx=\delta_{nn'},
\end{equation}
$n'=0,1,\ldots$, with $\delta_{nn'}$ being a Kronecker delta. In the limiting cases, the energy spectrum simplifies to:

at the small $\mathfrak{a}$:
\begin{align}\tag{17a$'$}\label{eq:17a'}
E_n(\mathfrak{a})&=\hbar\omega\left(2n+\frac{3}{2}-\mathfrak{a}^{1/2}+\mathfrak{a}+\ldots\right),\quad\mathfrak{a}\rightarrow0\\
\intertext{at the huge parameter:}
\tag{17a$''$}\label{eq:17a''}
E_n(\mathfrak{a})&=2\hbar\omega\left(n+\frac{1}{2}+\frac{1}{16\mathfrak{a}^{1/2}}-\frac{1}{256\mathfrak{a}^{3/2}}+\ldots\right),\quad\mathfrak{a}\rightarrow\infty.
\end{align}
Latter relation manifests that, as expected, the shift of the minimum to the right subdues the effect of the $x=0$ boundary what results in the gradual transformation of the potential to the HO shape with frequency $2\omega$. Observe that, according to equation~\eqref{ShcrodingerSol1_Energy}, the spectrum at any $\mathfrak{a}$ stays equidistant and the variation of this parameter from zero to the huge values shifts the energies downward by $\hbar\omega/2$ keeping the orbital-independent difference $\Delta E=E_{n+1}(\mathfrak{a})-E_n(\mathfrak{a})$ constant, $\Delta E=2\hbar\omega$. As it follows from equation~\eref{ShcrodingerSol1_Psi} and relation between Hermite $H_n(z)$ and Laguerre polynomials \cite{Abramowitz1}, at $\mathfrak{a}=0$ the spatial dependence reads \cite{Shi1}:
\begin{equation}\tag{17b$'$}\label{eq:17b'}
\Psi_n(0;x)=\frac{(-1)^n}{x_\omega^{1/2}}\frac{1}{2^{2n+\frac{1}{2}}}\left[\frac{1}{n!\Gamma\left(n+\frac{3}{2}\right)}\right]^\frac{1}{2}\exp\!\left(-\frac{1}{2}\frac{x^2}{x_\omega^2}\right)H_{2n+1}\!\left(\frac{x}{x_\omega}\right)
\end{equation}
and the corresponding eigen values are, according to equation~\eqref{eq:17a'}, those of the $\omega$-frequency HO with the odd indices $n$ only. Parts (b) and (c) of Figure~\ref{Fig_PotentialPositionWaveforms}, which depict waveforms of the ground- and first-excited orbitals, respectively, demonstrate that the increasing parameter $\mathfrak{a}$  transforms the function $\Psi_0(\mathfrak{a};x)$ [$\Psi_1(\mathfrak{a};x)$] into the dependence that becomes more and more symmetric (anti symmetric) with respect to $x_Z$. Let us also note that the divergence of the potential from equation~\eref{Potential1D_1} at the left edge $x=0$ forces the function to vanish there at any $\mathfrak{a}$ and $n$: $\Psi_n(\mathfrak{a};0)\equiv0$, what is directly seen from solutions~\eref{ShcrodingerSol1_Psi} and \eref{eq:17b'} and is depicted in panels (b) and (c).

\begin{figure}
\centering
\includegraphics[width=\columnwidth]{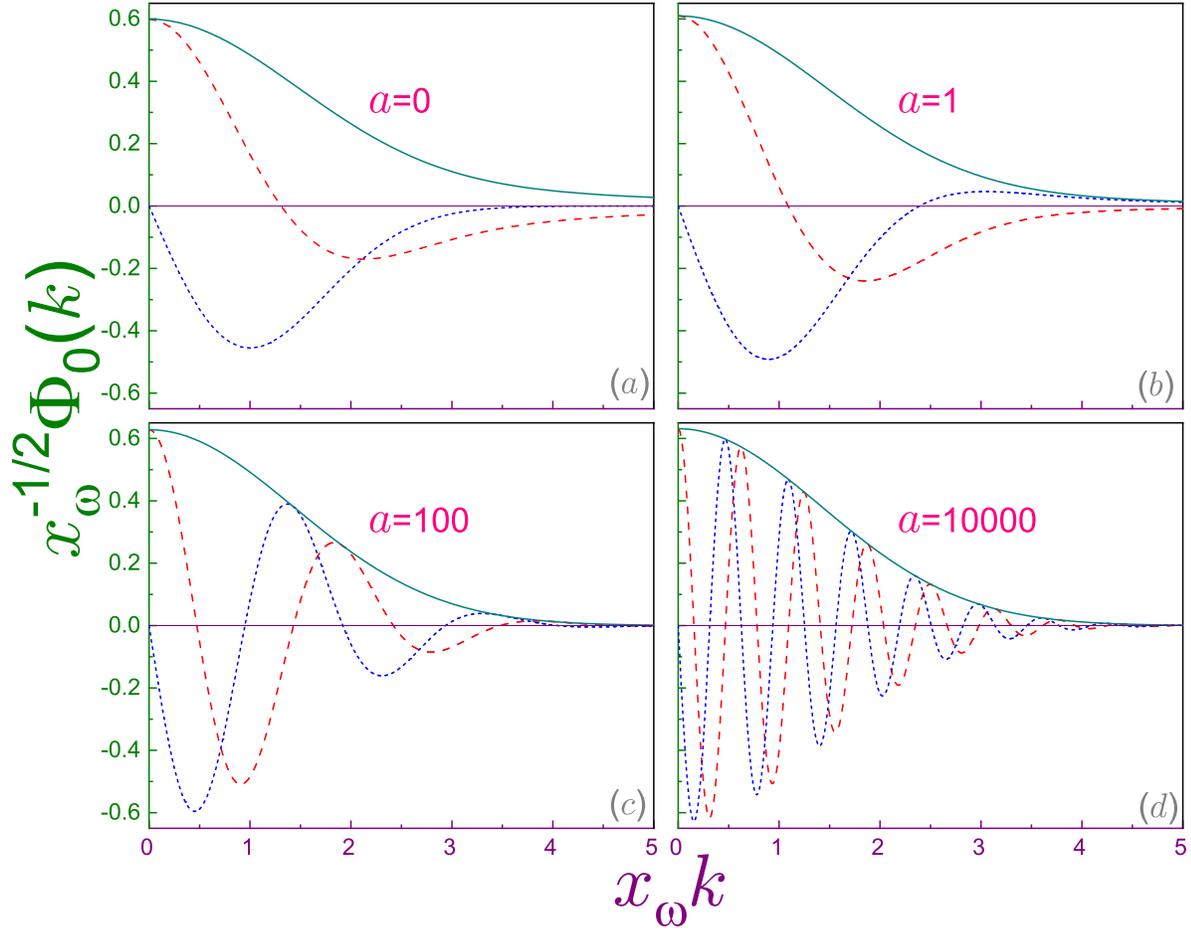}
\caption{\label{Fig_MomentumWaveform0}
Normalized ground-state momentum function in terms of the dimensionless wave vector $x_\omega k$ at several repulsive parameters: (a) $\mathfrak{a}=0$, (b) $\mathfrak{a}=1$, (c) $\mathfrak{a}=100$, (a) $\mathfrak{a}=10000$. Dashed (dotted) lines depict real (imaginary) parts of the waveform and the solid curve is its absolute value. Since real (imaginary) components are even (odd) functions of $k$, only the parts corresponding to the positive wave vectors are shown.}
\end{figure}

Momentum (or, more correctly, wave vector) dependencies $\Phi_n(k)$ are Fourier transforms of their position counterparts:
\begin{equation}\label{Fourier1}
\Phi_n(\mathfrak{a};k)=\frac{1}{(2\pi)^{1/2}}\int_0^\infty\Psi_n(\mathfrak{a};x)e^{-ikx}dx,
\end{equation}
and as such they inherit the orthonormalization from equation~\eqref{OrthoNormalization_Position1}:
\begin{equation}\label{OrthoNormalization_Momentum1}
\int_{-\infty}^\infty\Phi_n^*(\mathfrak{a};k)\Phi_{n'}(\mathfrak{a};k)dk=\delta_{nn'}.
\end{equation}
Fourier transform that is inverse to equation~\eref{Fourier1} reads:
\begin{equation}\label{Fourier2}
\Psi_n(\mathfrak{a};x)=\frac{1}{(2\pi)^{1/2}}\int_{-\infty}^\infty\Phi_n(\mathfrak{a};k)e^{ixk}dk.
\end{equation}

\begin{figure}
\centering
\includegraphics[width=\columnwidth]{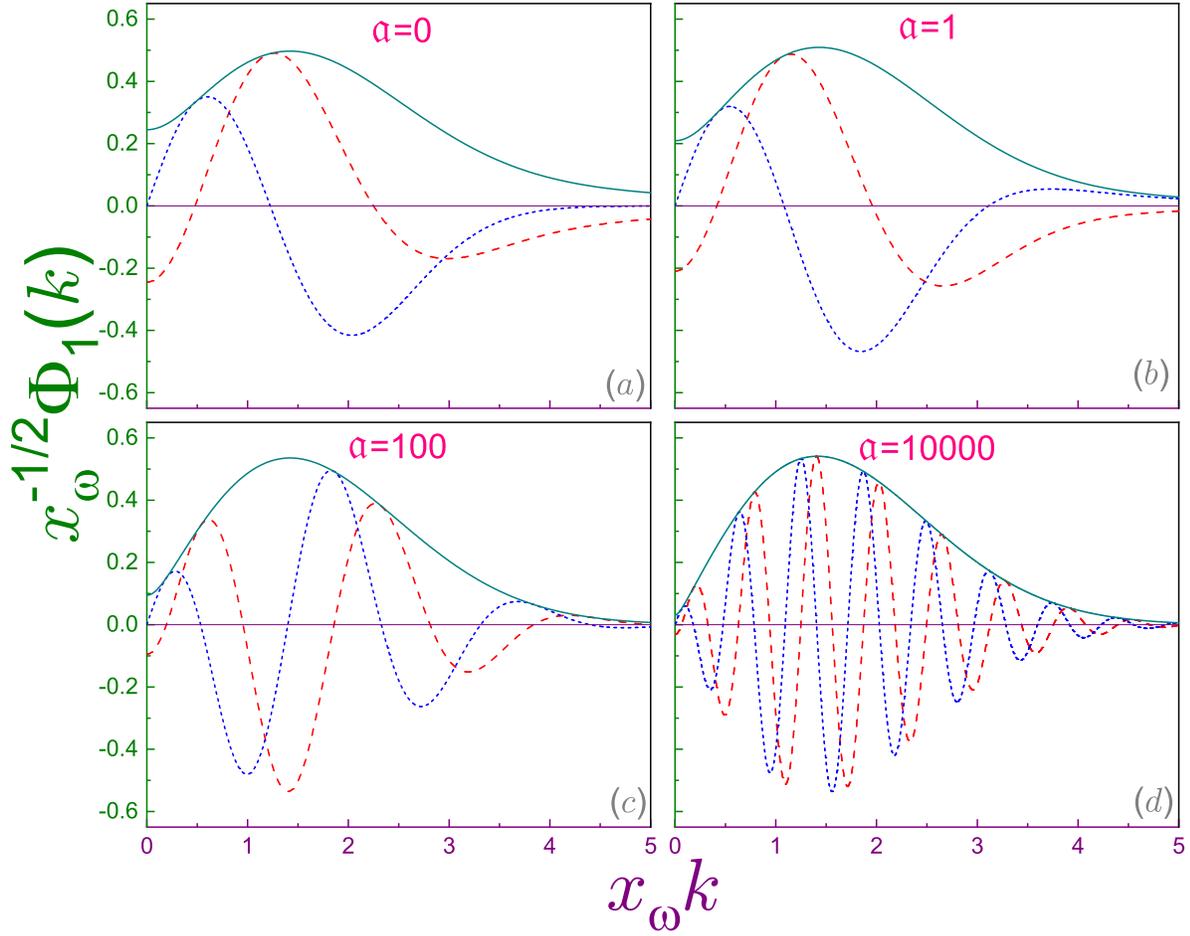}
\caption{\label{Fig_MomentumWaveform1}
The same as in Figure~\ref{Fig_MomentumWaveform0} but for the first excited state.}
\end{figure}

Substituting $\Psi_n(\mathfrak{a};x)$ from equation~\eref{ShcrodingerSol1_Psi} into \eref{Fourier1}, one gets after elementary algebra:
\begin{equation}\label{MomentumFunction1}
\Phi_n(\mathfrak{a};k)=x_\omega^{1/2}\left[\frac{n!}{\pi\Gamma(n+\eta+1)}\right]^\frac{1}{2}\int_0^\infty e^{-ikx_\omega y}e^{-y^2/2}y^{\eta+\frac{1}{2}}L_n^{(\eta)}\left(y^2\right)dy
\end{equation}
with the integral here being essentially complex:
\begin{equation}\label{Splitting1}
\Phi_n(k)=\Phi_n^{(r)}(k)+i\Phi_n^{(i)}(k),
\end{equation}
where both $\Phi_n^{(r)}(k)$ and $\Phi_n^{(i)}(k)$ are real and obey the even, $\Phi_n^{(r)}(-k)=\Phi_n^{(r)}(k)$, and odd, $\Phi_n^{(i)}(-k)=-\Phi_n^{(i)}(k)$, symmetry, respectively. Unfortunately, there is no in the known literature \cite{Abramowitz1,Gradshteyn1,Prudnikov1} any relevant formula for the integral in equation~\eref{MomentumFunction1}. To deal with $\Phi_n(\mathfrak{a};k)$ in our subsequent consideration, we found it convenient to use an explicit expression of the associated Laguerre polynomial in terms of powers of its argument \cite{Abramowitz1}:
\begin{equation}\label{Laguerre1}
L_n^{(\eta)}(z)=\sum_{j=0}^n(-1)^j\left(\begin{array}{c}
n+\eta\\
n-j
\end{array}\right)\frac{z^j}{j!},
\end{equation}
which, when substituted into equation~\eref{MomentumFunction1}, produces a finite sum of Weber parabolic cylinder functions $U(c,\xi)$ \cite{Abramowitz1} since their integral representation reads
\begin{equation}\label{Integral1}
\int_0^\infty e^{-ipy}e^{-y^2/2}y^{\eta+2j+\frac{1}{2}}dy=\Gamma\!\left(\eta+2j+\frac{1}{2}\right)e^{-p^2/4}U(\eta+2j,ip).
\end{equation}
Next, a splitting of the Weber function with purely imaginary argument into the real and imaginary parts \cite{Abramowitz1} leads to the same result of the momentum dependence, equation~\eref{Splitting1}; in particular, for the ground level one has:
\begin{subequations}\label{MomentumFunction0}
\begin{align}\label{MomentumFunction0_Real}
\Phi_0^{(r)}(\mathfrak{a};k)&=x_\omega^{1/2}\left[\frac{1}{\pi\Gamma(\eta+1)}\right]^\frac{1}{2}2^{\frac{\eta}{2}-\frac{1}{4}}\Gamma\!\left(\frac{\eta}{2}+\frac{3}{4}\right)e^{-\frac{x_\omega^2k^2}{2}}M\!\left(-\frac{\eta}{2}-\frac{1}{4},\frac{1}{2},\frac{x_\omega^2k^2}{2}\right)\\
\label{MomentumFunction0_Imaginary}
\Phi_0^{(i)}(\mathfrak{a};k)&=-x_\omega^{1/2}\!\left[\frac{1}{\pi\Gamma(\eta+1)}\right]^\frac{1}{2}2^{\frac{\eta}{2}+\frac{1}{4}}\Gamma\!\left(\frac{\eta}{2}+\frac{5}{4}\right)(x_\omega k)e^{-\frac{x_\omega^2k^2}{2}}M\!\!\left(-\frac{\eta}{2}+\frac{1}{4},\frac{3}{2},\frac{x_\omega^2k^2}{2}\right)
\end{align}
\end{subequations}
where Kummer confluent hypergeometric functions $M(p,q,z)$ \cite{Mathews1} emerge since the parabolic cylinder functions are expressed with their help \cite{Abramowitz1}. Figure~\ref{Fig_MomentumWaveform0} depicts dependencies from equations~\eref{MomentumFunction0} at several $\mathfrak{a}$. It is seen that the frequency of fading oscillations of the real $\Phi_0^{(r)}(\mathfrak{a};k)$ and imaginary $\Phi_0^{(i)}(\mathfrak{a};k)$ parts does increase at the growing parameter $\mathfrak{a}$ but, irrespective of it, the absolute value $|\Phi_0(\mathfrak{a};k)|$ is a smooth monotonically decreasing to zero function of the magnitude of the wave vector with no nodes on the $k$-axis. As was shown before \cite{Shi1}, the real part of the HHO momentum dependence of any orbital is expressed with the help of the imaginary error function ${\rm erfi}(z)=-i\,{\rm erf}(iz)$, with ${\rm erf}(z)$ being error function \cite{Abramowitz1} whereas $\Phi_n^{(i)}(0;k)$ is a polynomial of the degree $2n+1$ multiplied by $e^{-\frac{x_\omega^2k^2}{2}}$. In the opposite limit of the extremely huge $\mathfrak{a}$, the asymptotic behaviour of the Kummer functions \cite{Abramowitz1} yields:
\begin{equation}\label{MomentumFunction0_Asymptote1}
\Phi_0(\mathfrak{a};k)=\frac{x_\omega^{1/2}}{(2\pi)^{1/4}}\,e^{-\frac{x_\omega^2k^2}{4}}e^{-i2^{1/2}\mathfrak{a}^{1/4}x_\omega k}=\frac{x_{2\omega}^{1/2}}{\pi^{1/4}}\,e^{-\frac{x_{2\omega}^2k^2}{2}}e^{-i2\mathfrak{a}^{1/4}x_{2\omega}k},\quad\mathfrak{a}\rightarrow\infty,
\end{equation}
where the last step in this chain elementary follows, according to equation~\eref{Xomega}, from the identity $x_{2\omega}=x_\omega/2^{1/2}$. Second exponent in the right-hand side of relation~\eref{MomentumFunction0_Asymptote1} explains the increase of the frequency of the fading oscillations with the unrestricted growth of $\mathfrak{a}$ whereas all remaining multipliers there represent the ground-state momentum waveform \cite{Saha1,Olendski4,Griffiths1} of the HO with confining potential determined by $2\omega$. In a similar way, Kummer representation might be applied to the excited levels; for instance, 
\begin{subequations}\label{MomentumFunctionFirstExcited1}
\begin{align}
\Phi_1^{(r)}(\mathfrak{a};k)&=x_\omega^{1/2}\left[\frac{1}{\pi\Gamma(\eta+2)}\right]^\frac{1}{2}2^{\frac{\eta}{2}-\frac{5}{4}}\Gamma\!\left(\frac{\eta}{2}+\frac{3}{4}\right)e^{-\frac{x_\omega^2k^2}{2}}\nonumber\\
\label{MomentumFunctionFirstExcited1_Real}
&\times\left[2(1+\eta)M\!\left(-\frac{\eta}{2}-\frac{1}{4},\frac{1}{2},\frac{x_\omega^2k^2}{2}\right)-(3+2\eta)M\!\left(-\frac{\eta}{2}-\frac{5}{4},\frac{1}{2},\frac{x_\omega^2k^2}{2}\right)\right]\\
\Phi_1^{(i)}(\mathfrak{a};k)&=x_\omega^{1/2}\left[\frac{1}{\pi\Gamma(\eta+2)}\right]^\frac{1}{2}2^{\frac{\eta}{2}-\frac{3}{4}}\Gamma\!\left(\frac{\eta}{2}+\frac{5}{4}\right)(x_\omega k)e^{-\frac{x_\omega^2k^2}{2}}\nonumber\\
\label{MomentumFunctionFirstExcited1_Imaginary}
&\times\left[2(1+\eta)M\!\left(-\frac{\eta}{2}+\frac{1}{4},\frac{3}{2},\frac{x_\omega^2k^2}{2}\right)-(5+2\eta)M\!\left(-\frac{\eta}{2}-\frac{3}{4},\frac{3}{2},\frac{x_\omega^2k^2}{2}\right)\right].
\end{align}
\end{subequations}
Then, as Figure~\ref{Fig_MomentumWaveform1} exemplifies, the even function $|\Phi_1(\mathfrak{a};k)|$ exhibits two equal maxima located symmetrically with respect to the zero momentum whereas the $k=0$ minimum with the growth of the parameter $\mathfrak{a}$ decreases and at the huge $\mathfrak{a}$ is almost indistinguishable from zero with the whole $|\Phi_1(\mathfrak{a};k)|$  shape approaching closer and closer its $2\omega$ HO counterpart.

\begin{figure}
\centering
\includegraphics[width=\columnwidth]{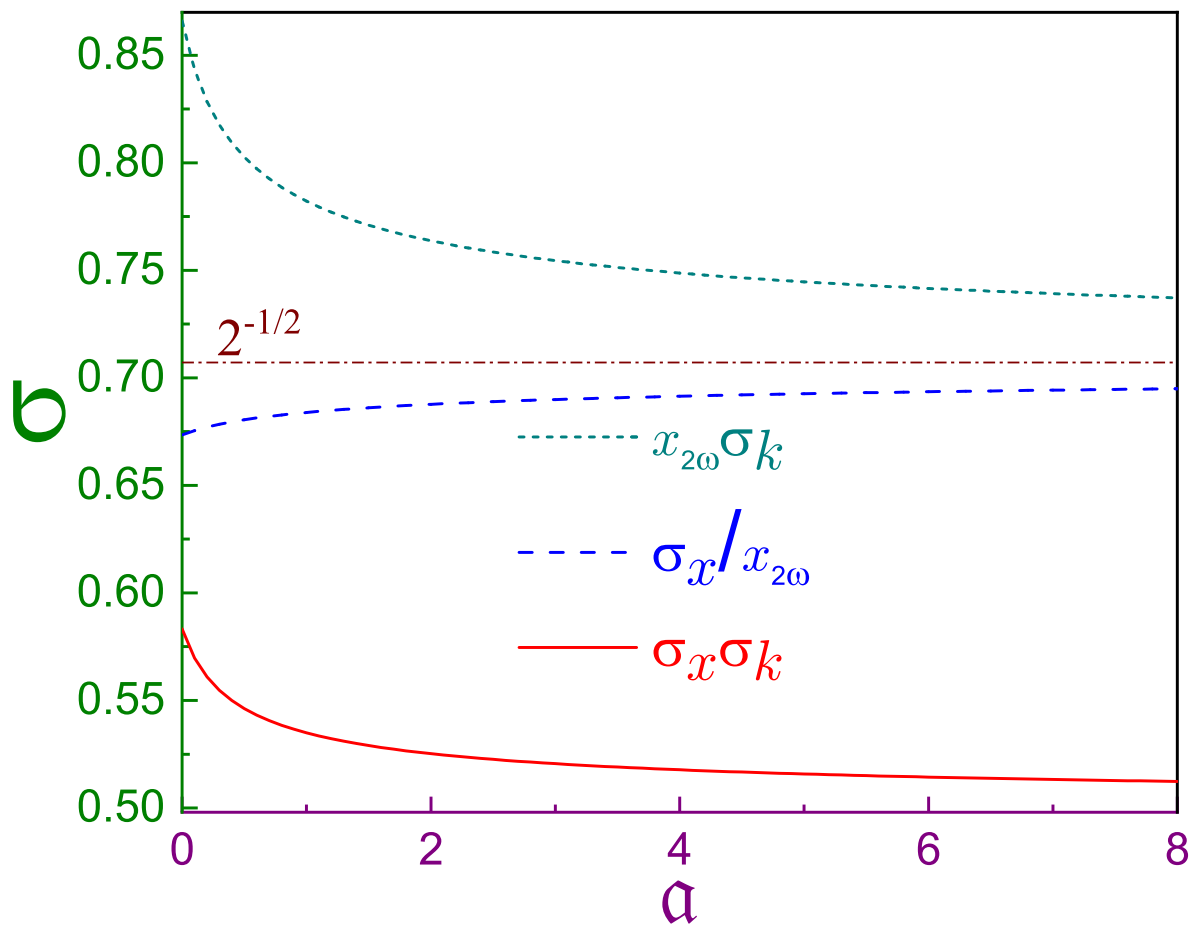}
\caption{\label{Fig_Heisenberg}
Dimensionless ground-state position $\sigma_x$ (dashed line, in units of $x_{2\omega}$) and wave vector  $\sigma_k$ (dotted curve, in units of $x_{2\omega}^{-1}$) standard deviations together with their product (solid dependence) in terms of the parameter $\mathfrak{a}$. Straight horizontal dash-dotted line depicts $2^{-1/2}=0.7071\ldots$.}
\end{figure}

\subsection{Standard deviations and Heisenberg uncertainty relation}\label{SubSec_Variance}
Understanding of the position and momentum waveforms developed in the previous subsection paves the way to the analysis of the miscellaneous measures describing different facets of the particle localization/delocalization, oscillating structure of its distribution and its deviation from the uniformity in the corresponding domain. The most famous of them that is known to each undergraduate student \cite{Griffiths1} is a standard deviation $\sigma$ that for the arbitrary 1D case has the following expression in the position and wave vector spaces:
\begin{subequations}\label{Variance1}
\begin{align}\label{Variance1_x}
\sigma_{x_n}&=\sqrt{\left\langle x^2\right\rangle_{x_n}-\left\langle x\right\rangle_{x_n}^2}\\
\label{Variance1_k}
\sigma_{k_n}&=\sqrt{\left\langle k^2\right\rangle_{k_n}-\left\langle k\right\rangle_{k_n}^2},
\end{align}
\end{subequations}
where the brackets $\langle\ldots\rangle_{x_n,k_n}$ denote quantum expectation values of the operators $\widehat{X}(x)$ or $\widehat{K}(k)$ for the orbital $n$ with subscripts denoting the domain over which this averaging is calculated:
\begin{subequations}\label{Expectation1}
\begin{align}\label{Expectation1_x}
\left\langle\widehat{X}\right\rangle_{x_n}&=\int_0^\infty\Psi_n^*(x)\widehat{X}(x)\Psi_n(x)dx\\
\label{Expectation1_k}
\left\langle\widehat{K}\right\rangle_{k_n}&=\int_{-\infty}^\infty\Phi_n^*(k)\widehat{K}(k)\Phi_n(k)dk.
\end{align}
\end{subequations}
If the corresponding operators simplify to $c$-numbers, i.e., $\widehat{X}(x)\equiv X(x)$ and $\widehat{K}(k)\equiv K(k)$, then expressions~\eref{Expectation1} read:
\begin{subequations}\label{Expectation2}
\begin{align}\label{Expectation2_x}
\left\langle X\right\rangle_{x_n}&=\int_0^\infty X(x)\rho_n(x)dx\\
\label{Expectation2_k}
\left\langle K\right\rangle_{k_n}&=\int_{-\infty}^\infty K(k)\gamma_n(k)dk,
\end{align}
\end{subequations}
$\rho_n(x)$ and $\gamma_n(k)$ are the distribution densities:
\begin{subequations}\label{Density1}
\begin{align}\label{Density1_x}
\rho_n(x)&=|\Psi_n(x)|^2\\
\label{Density1_k}
\gamma_n(k)&=\left|\Phi_n(k)\right|^2.
\end{align}
\end{subequations}
Below, if the averaging domain is obvious, the subscript $x$ or $k$ will be dropped.

Standard deviations are quantitative measures of indeterminacy of the particle behaviour and as such describe the amount of information about its location ($\sigma_x$) or motion ($\sigma_k$) that is \textit{not} known.  The product of two satisfies the following inequality:
\begin{equation}\label{Heisenberg1}
\sigma_{x_n}\sigma_{k_n}\geq\frac{1}{2},
\end{equation}
what is a mathematical expression of the celebrated Heisenberg uncertainty relation. It saturates for the HO ground state when the corresponding waveforms in either space are described by the Gaussian dependence \cite{Griffiths1}. Let us note that position (wave vector) component is measured in the units of length (inverse length) what makes equation~\eref{Heisenberg1} the most symmetric (and, accordingly, advantageous) compared to the use of the momentum ${\bf p}=\hbar{\bf k}$ instead of $\bf k$.

For PHO, the position mean value reads:
\begin{equation}\label{MeanValue1}
\langle x\rangle_n(\mathfrak{a})=\frac{3}{4}\sqrt{\frac{2}{\pi}}\frac{\Gamma\left(\eta+\frac{3}{2}\right)\Gamma\!\left(n-\frac{3}{2}\right)}{n!\,\Gamma(\eta+1)}\,{}_3F_2\!\left(-n,\eta+\frac{3}{2},\frac{3}{2};\eta+1,\frac{3}{2}-n;1\right)x_{2\omega}.
\end{equation}
Here, $_3F_2(a_1,a_2,a_3;b_1,b_2;z)$ is a generalized hypergeometric function \cite{Bailey1}. In arriving at this relation, the value of the integral \cite{Gradshteyn1,Prudnikov1}
$$\int_0^\infty e^{-z}z^{\eta+\frac{1}{2}}\left[L_n^{(\eta)}(z)\right]^2dz$$
has been used. It is worthwhile to point out that in equation~\eref{MeanValue1} and majority of all other relations below the lengths are expressed in units of $x_{2\omega}$. In this way, at the huge $\mathfrak{a}$ the quantities take the values characteristic to $2\omega$ HO. For the ground state, the above identity simplifies to
\begin{subequations}\label{MeanValue0}
\begin{align}\label{MeanValue0_1}
\langle x\rangle_0(\mathfrak{a})&=2^{1/2}\frac{\Gamma\left(\eta+\frac{3}{2}\right)}{\Gamma(\eta+1)}x_{2\omega}
\intertext{and its asymptotic cases are:}
\langle x\rangle_0(\mathfrak{a})&=2\sqrt{\frac{2}{\pi}}\left[1-\left(1-\frac{\ln2}{4}\right)\mathfrak{a}\right.\nonumber\\
\label{MeanValue0_0}
&\left.+\left(-\frac{\pi^2}{6}+3+2\ln^22-4\ln2\right)\mathfrak{a}^2+\ldots\right]x_{2\omega},\quad\mathfrak{a}\rightarrow0\\
\label{MeanValue0_Infinity}
\langle x\rangle_0(\mathfrak{a})&=\left(1+\frac{3}{8\mathfrak{a}^{1/2}}+\frac{1}{128\mathfrak{a}}+\ldots\right)x_Z,\quad\mathfrak{a}\rightarrow\infty.
\end{align}
\end{subequations}
Last equation shows that in the corresponding limit the function is localized around the minimum of the potential, as expected.

Expression $\langle x^2\rangle_n(\mathfrak{a})$ takes a simple form for any orbital:
\begin{equation}\label{MeanValueXsquare1}
\langle x^2\rangle_n(\mathfrak{a})=2(2n+\eta+1)x_{2\omega}^2.
\end{equation}
Accordingly, the ground-state position variance is calculated as
\begin{subequations}\label{PositionVariance1}
\begin{align}\label{PositionVariance1_1}
\sigma_{x_0}^2(\mathfrak{a})&=2\left[\eta+1-\frac{\Gamma^2\!\left(\eta+\frac{3}{2}\right)}{\Gamma^2(\eta+1)}\right]x_{2\omega}^2
\intertext{with asymptotes}\label{PositionVariance1_0}
\sigma_{x_0}^2(\mathfrak{a})&=\left[3-\frac{8}{\pi}+\left(2+\frac{16-32\ln2}{\pi}\right)\mathfrak{a}+\ldots\right]x_{2\omega}^2,\quad\mathfrak{a}\rightarrow0\\\label{PositionVariance1_Infinity}
\sigma_{x_0}^2(\mathfrak{a})&=\frac{1}{2}\left(1-\frac{1}{8\mathfrak{a}^{1/2}}+\frac{3}{64\mathfrak{a}}+\ldots\right)x_{2\omega}^2,\quad\mathfrak{a}\rightarrow\infty.
\end{align}
\end{subequations}
Here, $3-\frac{8}{\pi}=0.4535\ldots$ and its square root that will be used below is $\sqrt{3-\frac{8}{\pi}}=0.6734\ldots$.

In evaluating momentum variances and standard deviations, it has to be noticed first that, obviously, $\langle k\rangle_{k_n}=0$. For finding $\langle k^2\rangle_{k_n}$, one initially transforms it to averaging over the position space \cite{Griffiths1}:
\begin{equation}\label{Transform1}
\langle k^2\rangle_{k_n}=\left\langle-\frac{d^2}{dx^2}\right\rangle_{x_n}
\end{equation}
with $\hat{k}\equiv-i\frac{d}{dx}$ being a wave vector operator in $x$-representation. Next, multiplying Schr\"{o}dinger equation~\eref{Schrodinger1} by $\Psi_n^*(x)$ and integrating, the following relation is obtained:
\begin{equation}\label{Relation1}
\left\langle-\frac{d^2}{dx^2}\right\rangle_{x_n}=\frac{2m}{\hbar^2}\left(E_n-\left\langle V^{1D}\right\rangle_{x_n}\right).
\end{equation}
Here, 
\begin{equation}\label{Relation2}
\left\langle V^{1D}\right\rangle_{x_n}=\frac{1}{2}\hbar\omega\left(\frac{1}{x_\omega^2}\langle x^2\rangle_n-2\mathfrak{a}^{1/2}+\mathfrak{a}x_\omega^2\left\langle\frac{1}{x^2}\right\rangle_n\right).
\end{equation}
Last expectation value in the right-hand side is calculated analytically \cite{Prudnikov1}:
\begin{equation}\label{Relation3}
\left\langle\frac{1}{x^2}\right\rangle_n=\frac{1}{2\eta x_{2\omega}^2}.
\end{equation}
Ultimately,
\begin{subequations}\label{MomentumVariance1}
\begin{align}\label{MomentumVariance1_1}
\sigma_{k_n}^2(\mathfrak{a})&=\frac{2n+1+\frac{1}{2\sqrt{1+4\mathfrak{a}}}}{2x_{2\omega}^2}
\intertext{with asymptotes:}
\label{MomentumVariance1_0}
\sigma_{k_n}^2(\mathfrak{a})&=\frac{1}{2x_{2\omega}^2}\left(2n+\frac{3}{2}-\mathfrak{a}+3\mathfrak{a}^2-\ldots\right),\,\mathfrak{a}\rightarrow0\\
\label{MomentumVariance1_Infinity}
\sigma_{k_n}^2(\mathfrak{a})&=\frac{1}{2x_{2\omega}^2}\left(2n+1+\frac{1}{4\mathfrak{a}^{1/2}}-\frac{1}{32\mathfrak{a}^{3/2}}+\ldots\right),\,\mathfrak{a}\rightarrow\infty.
\end{align}
\end{subequations}
Then,
\begin{equation}\label{HeisenbergLimit1}
\sigma_{x_0}\sigma_{k_0}=\left\{\begin{array}{cc}
\frac{3^{1/2}}{2}\left(3-\frac{8}{\pi}\right)^{1/2}\left[1-\left(\frac{1}{3}-\frac{\pi+8-16\ln2}{3\pi-8}\right)\mathfrak{a}+\ldots\right],&\mathfrak{a}\rightarrow0\\
\frac{1}{2}\left(1+\frac{1}{16\mathfrak{a}^{1/2}}+\ldots\right)&\mathfrak{a}\rightarrow\infty
\end{array}
\right.,
\end{equation}
with $\frac{3^{1/2}}{2}\left(3-\frac{8}{\pi}\right)^{1/2}=0.5832\ldots$.

Figure~\ref{Fig_Heisenberg} shows evolution with $\mathfrak{a}$ of the ground-state dimensionless position $\sigma_{x_0}/x_{2\omega}$ and wave vector $x_{2\omega}\sigma_{k_0}$ standard deviations together with their product. Both parts are monotonic functions of the parameter $\mathfrak{a}$: position component grows from the above-mentioned HHO value of $\sqrt{3-\frac{8}{\pi}}$ to the inverse square root of two at $\mathfrak{a}\rightarrow\infty$ whereas its momentum counterpart decreases from $3^{1/2}/2=0.8660\ldots$ at $\mathfrak{a}=0$ to $2^{-1/2}=0.7071\ldots$ for $2\omega$ HO. This means that our knowledge about particle position (momentum) decreases (increases) with the enlarging $\mathfrak{a}$. Momentum rate of change dominates over the position variation; as a result, the left-hand side of the Heisenberg inequality~\eref{Heisenberg1} monotonically gets smaller from the above-mentioned $\sqrt{\frac{3}{2}\left(\frac{3}{2}-\frac{4}{\pi}\right)}$ to the fundamental limit of one half that consists of the two equal contributions from $\sigma_{x_0}$ and $\sigma_{k_0}$.

Let us finish this subsection by a brief note on the standard deviations of the excited states. It is known \cite{Griffiths1} that for the $2\omega$ HO their position and wave vector components are $\left(n+\frac{1}{2}\right)^{1/2}x_{2\omega}$ and $\left(n+\frac{1}{2}\right)^{1/2}x_{2\omega}^{-1}$, respectively, with their product $n+\frac{1}{2}$. Since at $n\geq1$ their variation with $\mathfrak{a}$ qualitatively reproduces their ground-orbital counterparts [what for, e.g., $\sigma_{k_n}(\mathfrak{a})$ is directly seen from equations~\eref{MomentumVariance1}], we do not plot them here.
\begin{figure}
\centering
\includegraphics[width=\columnwidth]{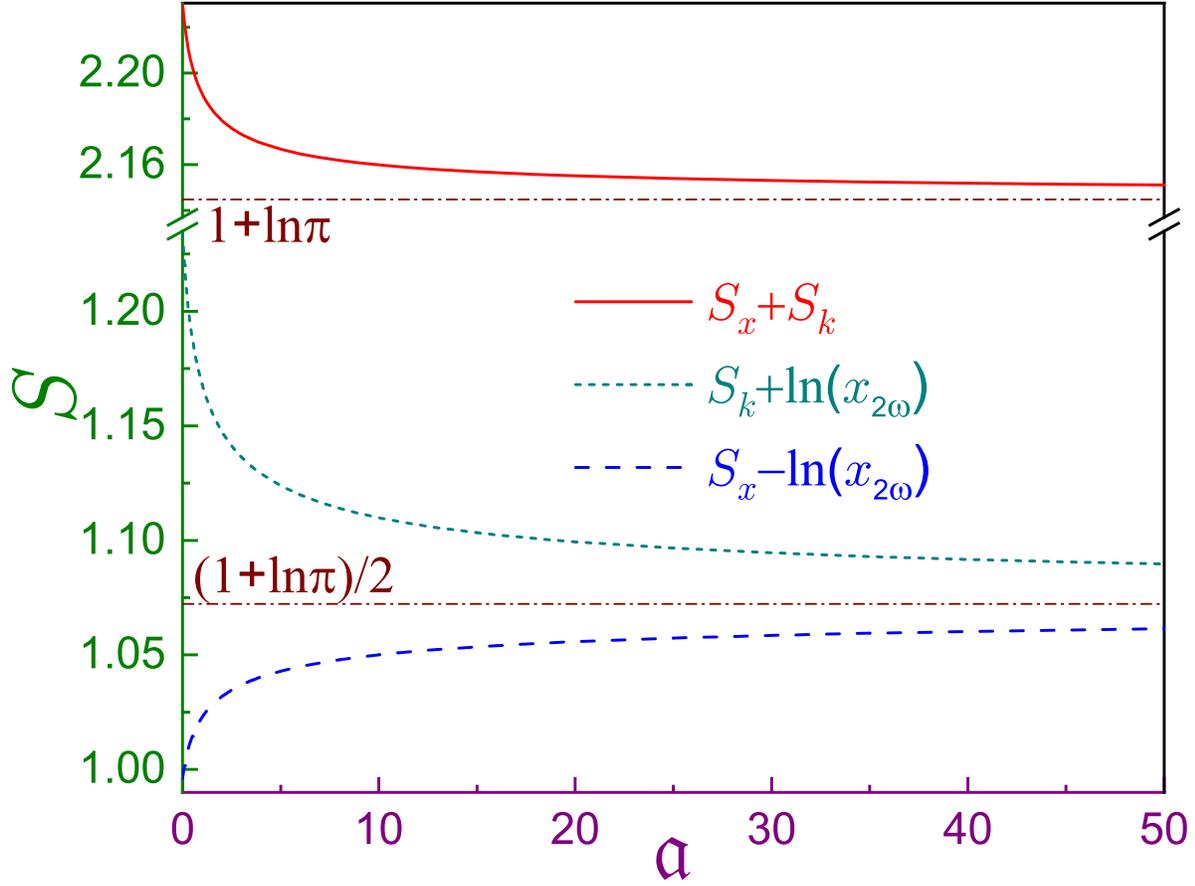}
\caption{\label{Fig_Shannon}
Dimensionless ground-state position $S_x-\ln x_{2\omega}$ (dashed line) and wave vector  $S_k+\ln x_{2\omega}$ (dotted curve) Shannon quantum information entropies together with their sum (solid dependence) in terms of the parameter $\mathfrak{a}$. Note vertical break from 1.235 to 2.137 and different scales below and above it. Lower and upper dash-dotted horizontal lines denote $(1+\ln\pi)/2=1.0723\ldots$ and $1+\ln\pi=2.1447\ldots$, respectively.}
\end{figure}
\subsection{Shannon quantum information entropies and non--Gaussianities}\label{SubSec_Shannon}
In general, for the motion in $\mathtt{d}$-dimensional space, Shannon position $S_{{\bf x}_\mathtt{n}}^{(\mathtt{d})}$ and wave vector $S_{{\bf k}_\mathtt{n}}^{(\mathtt{d})}$ entropies are defined as:
\begin{subequations}\label{Shannon1D}
\begin{align}\label{Shannon1D_R}
S_{{\bf x}_\mathtt{n}}^{(\mathtt{d})}&=-\int_{\mathcal{D}_{\bf x}^{(\mathtt{d})}}\rho_\mathtt{n}^{(\mathtt{d})}({\bf x})\ln\rho_{\mathtt{n}}^{(\mathtt{d})}({\bf x})d{\bf x}\\
\label{Shannon1D_K}
S_{{\bf k}_\mathtt{n}}^{(\mathtt{d})}&=-\int_{\mathcal{D}_{\bf k}^{(\mathtt{d})}}\gamma_\mathtt{n}^{(\mathtt{d})}({\bf k})\ln\gamma_\mathtt{n}^{(\mathtt{d})}({\bf k})d{\bf k},
\end{align}
\end{subequations} 
with $\mathcal{D}_{\bf x}^{(\mathtt{d})}$ and $\mathcal{D}_{\bf k}^{(\mathtt{d})}$  being the regions where the position $\rho_\mathtt{n}^{(\mathtt{d})}({\bf x})$ and momentum $\gamma_\mathtt{n}^{(\mathtt{d})}({\bf k})$ densities, respectively, are defined with hyperindex $\mathtt{n}$ counting the states in the increasing order of their energies. Two components of the Shannon quantum information entropy are not independent of each other but obey the inequality \cite{Bialynicki2,Beckner1}:
\begin{equation}\label{ShannonInequality1D}
S_{\bf x}^{(\mathtt{d})}+S_{\bf k}^{(\mathtt{d})}\geq\mathtt{d}(1+\ln\pi),
\end{equation}
with $1+\ln\pi=2.1447\ldots$. Similar to the quantities from the previous subsection, these measures quantify the amount of information that is \textit{not} known but, compared to $\sigma_{x,k}$, they present a much more general base for defining 'uncertainty' what, in particular, results in the fact that inequality~\eref{ShannonInequality1} is much stronger than its Heisenberg fellow \cite{Deutsch1}.

It is important to note that functionals from equations~\eref{Shannon1D} present an extension to the continuous distributions of the entropy of a discrete case:
\begin{equation}\label{ShannonDiscrete1}
S_p=-\sum_{n=1}^Np_n\ln p_n,
\end{equation}
as was done already by C. E. Shannon himself \cite{Shannon2}. Here, $0\leq p_n\leq1$, $n=1,2,\ldots,N$, is a probability belonging to a discrete set of all $N$ (that might by finite or infinite) possible events, so that $\sum_{n=1}^Np_n=1$. This has several implications. First, contrary to the \textit{dimensionless} $S_p$, position $S_{\bf x}^{(\mathtt{d})}$ and momentum $S_{\bf k}^{(\mathtt{d})}$ entropies are measured in units of, respectively, positive and negative logarithm of length times dimensionality $\mathtt{d}$. Second, discrete Shannon entropy is never negative varying between zero (sure event, one of $p_n$ is equal to unity with all others disappearing) and $\ln N$ (equiprobable situation, all happenings have the same chance to occur,  $p_n=1/N$) whereas its continuous counterparts might fall below zero what takes place when the parts of the corresponding density, say, $\rho_\mathtt{n}^{(\mathtt{d})}({\bf x})$, which are larger than unity, in their contribution to the integral in equation~\eref{Shannon1D_R} overweigh those segments where $\rho_\mathtt{n}^{(\mathtt{d})}({\bf x})<1$. In this regard, it is important and instructive to point out that in equations~\eref{Shannon1D} the classical rule from equation~\eref{ShannonDiscrete1} is applied to the position and momentum spreadings of the \textit{pure} state that is represented by the state vector. From this point of view, $S_{{\bf x}_\mathtt{n}}^{(\mathtt{d})}$ and $S_{{\bf k}_\mathtt{n}}^{(\mathtt{d})}$ should not be considered as 'quantum extensions' of the quantities defined for classical dependencies but rather they are 'classical quantities' applied to some quantum--related probability distributions. However, the most general quantum formulation employs the density operator $\widehat{\bm\rho}$ suitable for the description of the \textit{mixed} states. The corresponding von Neumann entropy
\begin{equation}\label{vonNeumann1}
\mathsf{S}\left(\widehat{\bm\rho}\right)=-{\rm Tr}\left(\widehat{\bm\rho}\ln\widehat{\bm\rho}\right)
\end{equation}
is non-negative and turns to zero for the pure state only. The distinction between quantum von Neumann entropy, equation~\eref{vonNeumann1}, and components of the Shannon quantum--information entropy, equations~\eref{Shannon1D}, applies equally well for the R\'{e}nyi and Tsallis measures considered in subsection~\ref{SubSec_Renyi}.

For the system under consideration, equations~\eref{Shannon1D} read:
\begin{subequations}\label{Shannon1}
\begin{align}\label{Shannon1_X}
S_{x_n}(\mathfrak{a})&=-\langle\ln\rho_n\rangle_{x_n}=-\int_0^\infty\rho_n(x)\ln\rho_n(x)dx\\
\label{Shannon1_K}
S_{k_n}(\mathfrak{a})&=-\langle\ln\gamma_n\rangle_{k_n}=-\int_{-\infty}^\infty\gamma_n(k)\ln\gamma_n(k)dk
\end{align}
\end{subequations}
and inequality~\eref{ShannonInequality1D} degenerates to
\begin{equation}\label{ShannonInequality1}
S_{x_n}+S_{k_n}\geq1+\ln\pi,
\end{equation}
that is transformed into the identity for the Gaussian distributions, as it was the case for the standard deviations too. For the HO with its characteristic length $x_\omega$, one has $S_{x_0}=\ln x_\omega+(1+\ln\pi)/2$ and $S_{k_0}=-\ln x_\omega+(1+\ln\pi)/2$.

As already stressed above, Shannon entropies of the continuous distributions from equations~\eref{Shannon1} are measured in units of the positive (position part) or negative (wave vector component) logarithm of the length. To avoid this physical ambiguity, in line with the previous discussions \cite{Olendski6,Olendski7}, below the dimensionless quantities $S_{x_n}-\ln x_{2\omega}$ and $S_{k_n}+\ln x_{2\omega}$ are studied.

Due to the complicated form  of the integrands in equations~\eref{Shannon1} for the excited orbitals, only ground-state position measure can be written analytically:
\begin{subequations}\label{PositionShannon0}
\begin{align}\label{PositionShannon0_1}
S_{x_0}(\mathfrak{a})-\ln x_{2\omega}&=\frac{1}{2}\ln2+\ln\frac{\Gamma(\eta+1)}{2}-\left(\eta+\frac{1}{2}\right)\psi(\eta)+\eta-\frac{1}{2\eta},
\intertext{$\psi(z)=\frac{\Gamma'(z)}{\Gamma(z)}$ is $\psi$-function \cite{Abramowitz1}. Limiting cases of equation~\eqref{PositionShannon0_1} are:}\label{PositionShannon0_0}
S_{x_0}(\mathfrak{a})-\ln x_{2\omega}&=\frac{1}{2}\ln(2\pi)-\frac{1}{2}+\gamma+\left(5-\frac{\pi^2}{2}\right)\mathfrak{a}+\ldots,\,\mathfrak{a}\rightarrow0\\
\label{PositionShannon0_Infinity}
S_{x_0}(\mathfrak{a})-\ln x_{2\omega}&=\frac{1}{2}(1+\ln\pi)-\frac{1}{12\mathfrak{a}^{1/2}}+\frac{1}{24\mathfrak{a}}+\ldots,\,\mathfrak{a}\rightarrow\infty.
\end{align}
\end{subequations}
Here, $\gamma$ is Euler's constant \cite{Abramowitz1}, $\frac{1}{2}\ln(2\pi)-\frac{1}{2}+\gamma=0.9961\ldots$ and $(1+\ln\pi)/2=1.0723\ldots$. Leading term of right-hand side of equation~\eref{PositionShannon0_0} does coincide with the HHO Shannon entropy \cite{Shi1}. All other position and momentum measures are calculated numerically only.

Ground-state dimensionless position $S_{x_0}(\mathfrak{a})-\ln x_{2\omega}$ and wave vector $S_{k_0}(\mathfrak{a})+\ln x_{2\omega}$ entropies together with their sum are shown in figure~\ref{Fig_Shannon}. The growth of $\mathfrak{a}$ decreases (increases) our knowledge about particle location (motion): position (wave vector) component monotonically gets bigger (smaller). Remarkably, the loss or gain of information is a purely quantum phenomenon: in classical picture, the diameter $x_+-x_-$ is not influenced by $\mathfrak{a}$, as it follows from equation~\eref{Diameter1}. Similar to the standard deviations, section~\ref{SubSec_Variance}, momentum rate of change prevails over its position counterpart: as the potential with the enlarging $\mathfrak{a}$ shapes into the $2\omega$ HO, the total \textit{un}available information shrinks to the fundamental limit of $1+\ln\pi$ approaching it from above with equal contribution from each item.

As was already stated several times, at large $\mathfrak{a}$ the ground orbital comes closer and closer to the Gaussian shape. To quantify the deviation from Gaussianity, we adapt for our geometry the measure introduced for the von Neumann entropy \cite{Genoni1}; namely, position $\delta_x^{(nG)}$ and momentum $\delta_k^{(nG)}$ non--Gaussianities for the PHO read:
\begin{subequations}\label{nonGaussianity1}
\begin{align}\label{nonGaussianity1_X}
\delta_x^{(nG)}&=\int_0^\infty dx\rho_0(x)\ln\frac{\rho_0(x)}{\tau_x(x)}\\
\label{nonGaussianity1_K}
\delta_k^{(nG)}&=\int_{-\infty}^\infty dk\gamma_0(k)\ln\frac{\gamma_0(k)}{\tau_k(k)}
\end{align}
\end{subequations}
which are nothing else but relative entropies or Kullback--Leibler divergences \cite{Kullback1}. Here, $\tau_x$ and $\tau_k$ are reference position and wave vector Gaussian states that have the same first and second position $\langle x\rangle_\tau$, $\langle x^2\rangle_\tau$ and wave vector $\langle k\rangle_\tau$, $\langle k^2\rangle_\tau$ moments, respectively, as the PHO ground level. Properties of the dimensionless functionals from equations~\eref{nonGaussianity1} are very similar to those of the mixed states \cite{Genoni1,Genoni2}; in particular,
\begin{equation}\label{nonGaussianity2}
\delta_{x,k}^{(nG)}=S_{\tau_x,\tau_k}-S_{x_0,k_0}.
\end{equation}
Their calculation is quite straightforward; for example, recalling that for the centered at the origin position HO with characteristic length $x_\tau$ the variance is $x_\tau^2/2$, one equates it with its counterpart from equation~\eref{PositionVariance1_1} finding in this way $x_\tau$. Its knowledge effortlessly leads to the evaluation of the associated position entropy $S_{\tau_x}$ yielding ultimately:
\begin{subequations}\label{nonGaussianityPosition1}
\begin{align}
\delta_x^{(nG)}(\mathfrak{a})&=\frac{1+\ln\pi}{2}+\frac{3}{2}\ln2+\frac{1}{2}\ln\!\left(\!\eta+\frac{\Gamma^2\left(\eta+\frac{3}{2}\right)}{\Gamma^2\left(\eta+1\right)}-\!1\!\right)\nonumber\\\label{nonGaussianityPosition1_1}
&-\ln\Gamma(\eta+1)+\left(\eta+\frac{1}{2}\right)\psi(\eta)-\eta+\frac{1}{2\eta}
\intertext{with the asymptotes:}
\label{nonGaussianityPosition1_0}
\delta_x^{(nG)}(\mathfrak{a})&=\frac{1}{2}\ln(3\pi-8)\!-\!\frac{1}{2}\ln\pi\!+\!1\!-\!\gamma\!+\!\frac{3\pi^3\!-\!8\pi^2\!-\!28\pi\!+\!96\!-\!32\ln\!2}{6\pi-16}\mathfrak{a}+\ldots,\,\mathfrak{a}\rightarrow0\\
\label{nonGaussianityPosition1_Infinity}
\delta_x^{(nG)}(\mathfrak{a})&=\frac{1}{48\mathfrak{a}^{1/2}}-\frac{17}{768\mathfrak{a}}+\ldots,\,\mathfrak{a}\rightarrow\infty.
\end{align}
\end{subequations}
Here, $\frac{1}{2}\ln(3\pi-8)-\frac{1}{2}\ln\pi+1-\gamma=0.02742\ldots$ and $(3\pi^3-8\pi^2-28\pi+96-32\ln\!2)/{(6\pi-16)}=-0.02923\ldots$. The sign of the last number shows that the position non--Gaussianity decreases at the small $\mathfrak{a}$ and equation~\eref{nonGaussianityPosition1_Infinity} expands this statement to the huge parameter. In a similar way, one finds the characteristic length of the reference momentum Gaussian state, which in general is different from its position fellow, and then evaluates the associated entropy:
\begin{equation}
S_{\tau_k}(\mathfrak{a})=-\ln x_{2\omega}+\frac{1+\ln\pi}{2}+\frac{1}{2}\ln\!\left(1+\frac{1}{2\sqrt{1+4\mathfrak{a}}}\right).
\end{equation}
\begin{figure}
\centering
\includegraphics[width=\columnwidth]{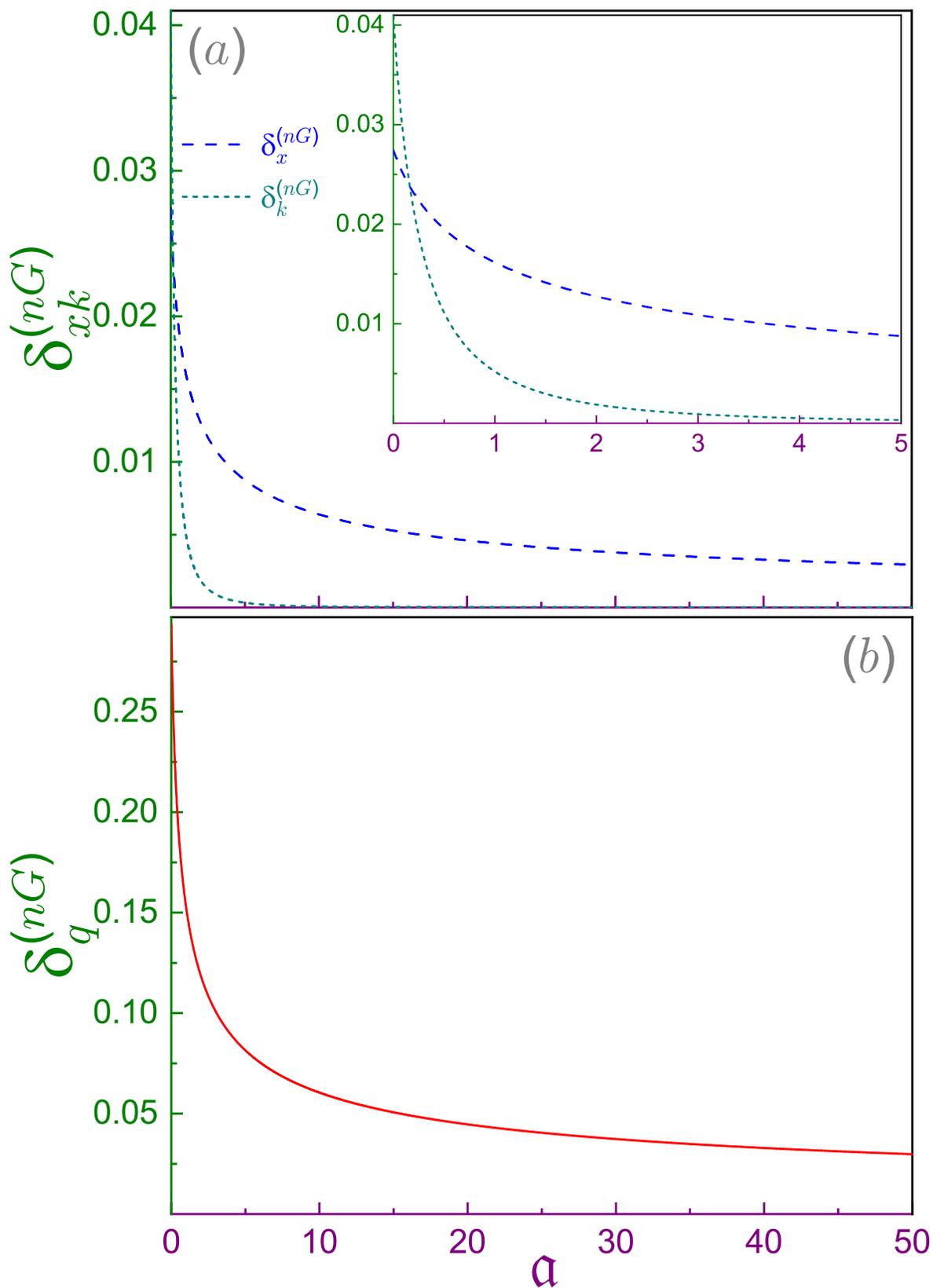}
\caption{\label{Fig_nonGaussianity}
(a) Position $\delta_x^{(nG)}$ (dashed line) and momentum $\delta_k^{(nG)}$ (dotted curve) non--Gaussianities as functions of parameter $\mathfrak{a}$. Inset  shows an enlarged view at $0\leq\mathfrak{a}\leq5$. (b) Measure $\delta_q^{(nG)}$ as a function of $\mathfrak{a}$. Note different vertical ranges in each window.}
\end{figure}

Figure~\ref{Fig_nonGaussianity}(a) shows evolution of both non--Gaussianities with the parameter $\mathfrak{a}$. Either of them is a positive monotonically decreasing function of $\mathfrak{a}$ that tends to zero in the limit $\mathfrak{a}\rightarrow\infty$. For the HHO, momentum component with its value of $0.04095\ldots$ is greater than its position counterpart but the magnitude of its speed of change at the small $\mathfrak{a}$ is faster and at $\mathfrak{a}\gtrsim0.16$ it plunges below $\delta_x^{(nG)}$ and stays much closer to zero. Interestingly, such asymptotic behaviour is opposite to that of the Shannon entropies, as a comparison between figures~\ref{Fig_Shannon} and \ref{Fig_nonGaussianity} reveals.

Having seen the evolution of $\delta_x^{(nG)}(\mathfrak{a})$ and $\delta_k^{(nG)}(\mathfrak{a})$, which are 'classical' measures of the quantum probability position and wave vector distributions, one might wonder how they match up with essentially quantum non--Gaussianity $\delta_q^{(nG)}(\mathfrak{a})$ that for the single-mode orbitals reads \cite{Paris1}:
\begin{subequations}\label{nonGaussianityQuantum1}
\begin{align}\label{nonGaussianityQuantum1_A}
\delta_q^{(nG)}&=h\!\left(\!\sqrt{\det({\bm\sigma}_{cov})}\right)-\mathsf{S}\left(\widehat{\bm\rho}\right),
\intertext{and since for the pure states, which we only consider here, $\mathsf{S}\left(\widehat{\bm\rho}\right)=0$, this relation simplifies to:}
\label{nonGaussianityQuantum1_B}
\delta_q^{(nG)}&=h\!\left(\!\sqrt{\det({\bm\sigma}_{cov})}\right)
\end{align}
\end{subequations}
with
\begin{equation}\label{h1}
h(x)=\!\left(\!x+\frac{1}{2}\right)\ln\!\left(\!x+\frac{1}{2}\right)-\!\left(\!x-\frac{1}{2}\right)\ln\!\left(\!x-\frac{1}{2}\right).
\end{equation}
Dimensionless elements of the (symmetric) covariance matrix
\begin{subequations}
\begin{align}
{\bm\sigma}_{cov}&=\left(
\begin{array}{cc}
\sigma_{11}&\sigma_{12}\\
\sigma_{21}&\sigma_{22}
\end{array}
\right)
\intertext{are:}
\sigma_{11}&=\frac{\sigma_{x_0}^2}{x_{2\omega}^2}\\
\sigma_{22}&=x_{2\omega}^2\sigma_{k_0}^2\\
\sigma_{12}&=\sigma_{21}=\frac{1}{2}\left\langle\hat{x}\hat{k}+\hat{k}\hat{x}\right\rangle_{x_0}-\langle x\rangle_0\langle k\rangle_0.
\end{align}
\end{subequations}
Similar to other 1D structures considered before \cite{Paris1}, it is elementary to show that there are no correlations for our potential either what makes the covariance matrix a diagonal one, $\sigma_{12}=\sigma_{21}=0$.

Panel (b) of figure~\ref{Fig_nonGaussianity} depicts the change of $\delta_q^{(nG)}$ with $\mathfrak{a}$. Similar to its position and momentum counterparts discussed above, it is a monotonically decreasing function of its argument being however, a few times greater than them; for example, close to the HHO, it behaves as
\begin{subequations}\label{nonGaussianityQuantum2}
\begin{align}\label{nonGaussianityQuantum2_0}
\delta_q^{(nG)}(\mathfrak{a})&=\!h\!\left(\frac{3^{1/2}}{2}\sqrt{\!3\!-\!\frac{8}{\pi}}\right)\!\!+\!\frac{8}{\pi}\sqrt{\frac{3}{3\!-\!\frac{8}{\pi}}}\!\left(\!\ln2\!-\!\frac{2}{3}\right)\!\ln\!\!\left(\!\frac{\sqrt{3\!\left(3\!-\!\frac{8}{\pi}\right)}\!-\!1}{\sqrt{3\left(3\!-\!\frac{8}{\pi}\right)}\!+\!1}\right)\!\!\mathfrak{a}\!+\!\ldots,\,\mathfrak{a}\rightarrow0,
\intertext{$h\!\left(\frac{3^{1/2}}{2}\sqrt{3\!-\!\frac{8}{\pi}}\right)=0.2934\ldots$ and the coefficient at $\mathfrak{a}$ being negative: $-0.4450\ldots$. In the opposite regime when the potential shape approaches the perfectly symmetric double frequency oscillator, the non--Gaussianity fades to zero according to:}
\label{nonGaussianityQuantum2_Infinity}
\delta_q^{(nG)}(\mathfrak{a})&=\frac{1}{32\mathfrak{a}^{1/2}}\left(\frac{1}{2}\ln\mathfrak{a}+5\ln2+1\right)+\ldots,\,\mathfrak{a}\rightarrow\infty.
\end{align}
\end{subequations}
Comparison of the latter relation with the position asympote from equation~\eref{nonGaussianityPosition1_Infinity} shows much slower wane of $\delta_q^{(nG)}$.

\begin{figure}
\centering
\includegraphics[width=\columnwidth]{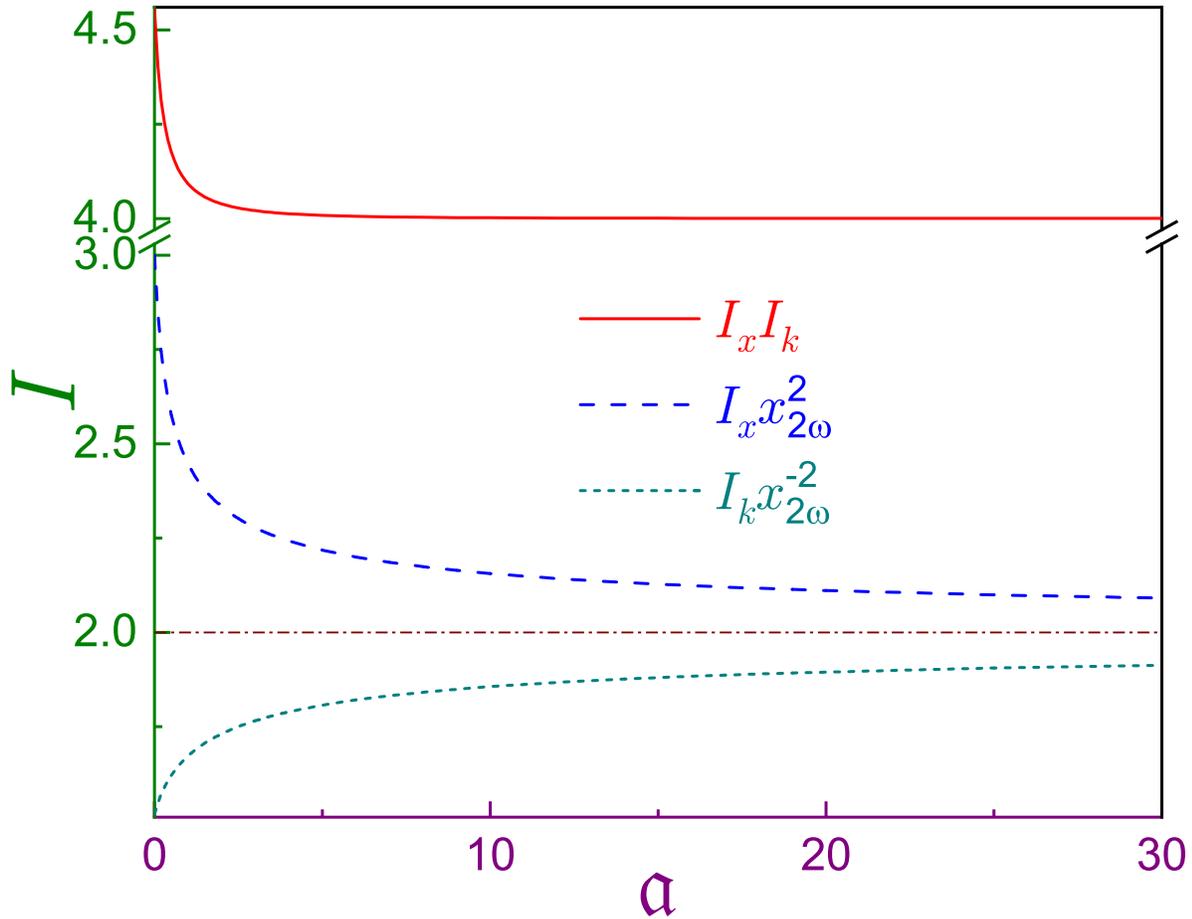}
\caption{\label{Fig_Fisher}
Dimensionless ground-state position $x_{2\omega}^2I_x$ (dashed line) and wave vector  $x_{2\omega}^{-2}I_k$ (dotted curve) Fisher informations together with their product (solid dependence) in terms of the parameter $\mathfrak{a}$. Note vertical break from 3.03 to 3.97. Dash-dotted horizontal line denotes the value of 2.}
\end{figure}
\subsection{Fisher informations}\label{SubSec_Fisher}
Whereas Shannon entropy is a global measure of uncertainty, presence of the gradients ${\bm\nabla}_{\bf x}$ and ${\bm\nabla}_{\bf k}$ in the mathematical expressions of Fisher informations \cite{Fisher1} makes them local indicators appraising the rates of change of the corresponding distribution:
\begin{subequations}\label{Fisher1}
\begin{align}\label{Fisher1_R}
I_{{\bf x}_\mathtt{n}}^{(\mathtt{d})}&=\int_{\mathcal{D}_\rho^{(\mathtt{d})}}\rho_\mathtt{n}^{(\mathtt{d})}({\bf x})\left|{\bm\nabla}_{\bf x}\ln\rho_\mathtt{n}^{(\mathtt{d})}({\bf x})\right|^2\!\!d{\bf x}=\int_{\mathcal{D}_\rho^{(\mathtt{d})}}\frac{\left|{\bm\nabla}_{\bf x}\rho_\mathtt{n}^{(\mathtt{d})}({\bf x})\right|^2}{\rho_\mathtt{n}^{(\mathtt{d})}({\bf x})}d{\bf x}\\
\label{Fisher1_K}
I_{{\bf k}_\mathtt{n}}^{(\mathtt{d})}&=\int_{\mathcal{D}_\gamma^{(\mathtt{d})}}\gamma_\mathtt{n}^{(\mathtt{d})}({\bf k})\left|{\bm\nabla}_{\bf k}\ln\gamma_\mathtt{n}^{(\mathtt{d})}({\bf k})\right|^2\!\!d{\bf k}=\int_{\mathcal{D}_\gamma^{(\mathtt{d})}}\frac{\left|{\bm\nabla}_{\bf k}\gamma_\mathtt{n}^{(\mathtt{d})}({\bf k})\right|^2}{\gamma_\mathtt{n}^{(\mathtt{d})}({\bf k})}\,d{\bf k}.
\end{align}
\end{subequations}
These equations state that the steeper or more oscillatory the distribution is, the larger the corresponding Fisher term becomes. Avalanche-like growing diversity of applications of this quantity is reflected in a remarkable fact: first edition of book~\cite{Frieden1} was entitled 'Physics from Fisher information' and in the later printings it has been changed to its present wording. Important for our consideration, one has to point out at the crucial role of this measure in building the bridge between the energy and information:  it was shown that in density-functional theory its position component is nothing else than the functional of the kinetic energy of the many-particle system \cite{Sears1}. Contrary to the Shannon entropy, inequality~\eref{ShannonInequality1D}, there is no similar universal relation between position and momentum components of the Fisher information. For the $\mathtt{d}$D Gaussian distribution with characteristic frequency $\omega$, i.e., for the HO lowest-energy orbital, $\mathtt{n}=0$, dimensionless parts are equal to each other \cite{Dehesa2}, as is the case with all other measures too:
\begin{equation}\label{FisherHOD1}
x_\omega^2I_{\bf x_\mathtt{0}}^{(\mathtt{d})HO}=x_\omega^{-2}I_{\bf k_\mathtt{0}}^{(\mathtt{d})HO}=2\mathtt{d}.
\end{equation}

For the 1D PHO, equations~\eref{Fisher1} become:
\begin{subequations}\label{Fisher2}
\begin{align}\label{Fisher2_R}
I_{x_n}(\mathfrak{a})&=\int_0^\infty\frac{1}{\rho_n(x)}\left[\frac{d}{dx}\rho_n(x)\right]^2dx\\
\label{Fisher2_K}
I_{k_n}(\mathfrak{a})&=\int_{-\infty}^\infty\frac{1}{\gamma_n(k)}\left[\frac{d}{dk}\gamma_n(k)\right]^2dk.
\end{align}
\end{subequations}
After straightforward \cite{Gradshteyn1,Prudnikov1} but lengthy calculation, the position component takes the form:
\begin{subequations}\label{Fisher3}
\begin{align}\label{Fisher3_1}
I_{x_n}(\mathfrak{a})&=\frac{2}{x_{2\omega}^2}\left(2n+1+\frac{1}{4\eta}\right)
\intertext{with the asymptotes:}\label{Fisher3_0}
I_{x_n}(\mathfrak{a})&=\frac{2}{x_{2\omega}^2}\left(2n+\frac{3}{2}-\mathfrak{a}+3\mathfrak{a}^2-\ldots\right),\quad\mathfrak{a}\rightarrow0\\
\label{Fisher3_Infinity}
I_{x_n}(\mathfrak{a})&=\frac{2}{x_{2\omega}^2}\left(2n+1+\frac{1}{4\mathfrak{a}^{1/2}}-\frac{1}{32\mathfrak{a}^{3/2}}+\ldots\right),\quad\mathfrak{a}\rightarrow\infty.
\end{align}
\end{subequations}
In equation~\eref{Fisher3_0}, the leading term describes a contribution from the HHO \cite{Shi1}.

Energy spectrum, equation~\eref{ShcrodingerSol1_Energy}, and $I_x-\mathfrak{a}$ characteristics~\eref{Fisher3_1} have many features in common: both are equidistant with linear dependence on quantum index, both are monotonically decreasing functions with almost coinciding shapes what can be, for example, seen from the same asymptotic behaviour at huge $\mathfrak{a}$, cf. equations~\eref{eq:17a''} and \eref{Fisher3_Infinity}. This very close similarity is another exemplification of the intimate relation between the energy and position Fisher information whose $n=0$ curve is shown in figure~\ref{Fig_Fisher}: with the the increase of the parameter $\mathfrak{a}$, from its $\mathfrak{a}=0$ value of $3$ it steadily approaches from above $2\omega$ HO magnitude of two.

Momentum components are calculated numerically only. Dotted curve in figure~\ref{Fig_Fisher} shows that its $n=0$ dependence grows with $\mathfrak{a}$ (as all other $I_{k_{n\geq1}}$ do too): for HHO, it is equal to $1.518\ldots$ whereas in the opposite regime it nears the value of the HO with frequency $2\omega$. Similar to its position counterpart, $I_{k_n}$ at any $\mathfrak{a}$ is an increasing function of the index. Note that at $\mathfrak{a}\gtrsim5$, position and wave vector parts are almost vertically symmetric with respect to $I=2$ line. As a result, their product in this range is practically indistinguishable from four, e. g., $I_{x_0}(5)I_{k_0}(5)=4.00843\ldots$. For all other quantities studied here, the convergence of the combined measure is much slower.

\begin{figure}
\centering
\includegraphics[width=\columnwidth]{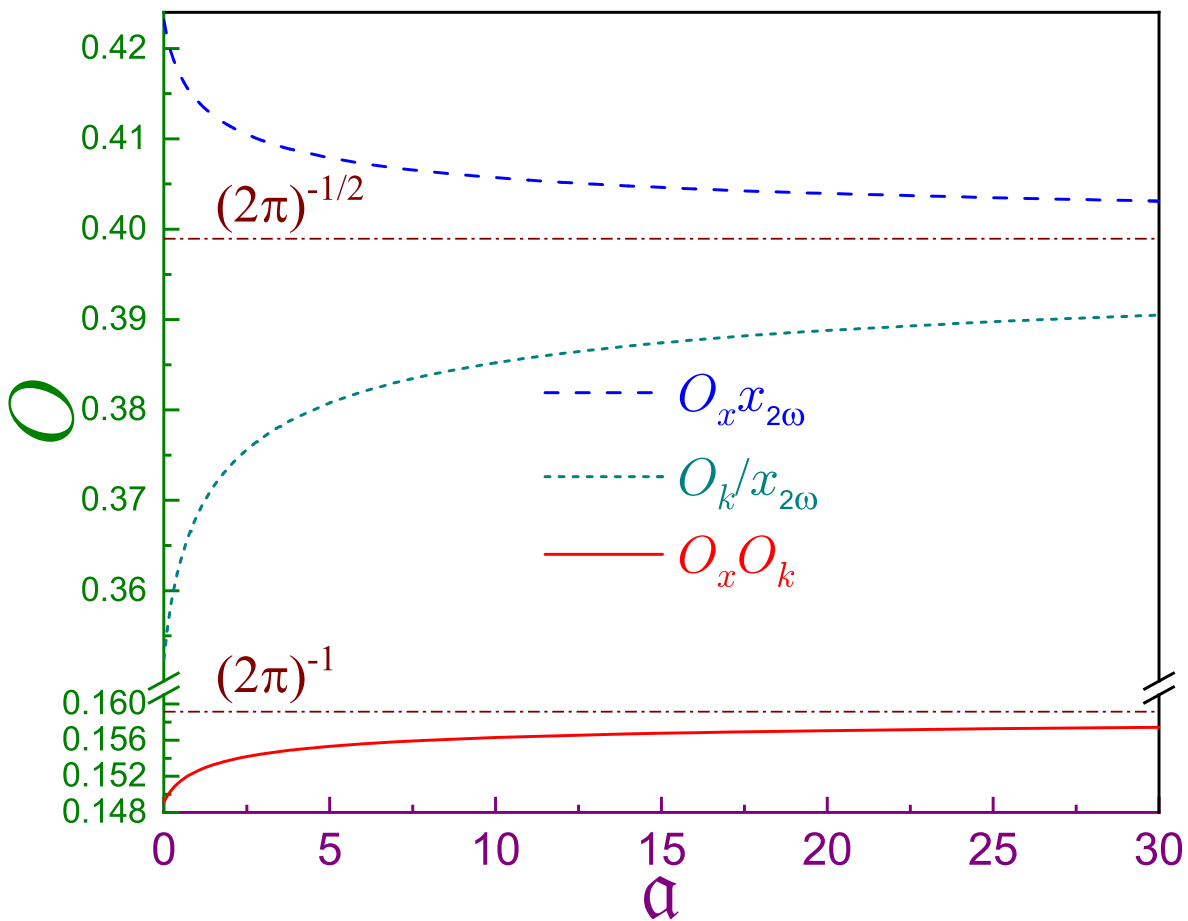}
\caption{\label{Fig_Onicescu}
Dimensionless ground-state position $O_xx_{2\omega}$ (dashed line) and wave vector  $O_k/x_{2\omega}$ (dotted curve) Onicescu energies together with their product (solid dependence) in terms of the parameter $\mathfrak{a}$. Note vertical break from 0.161 to 0.35 and different scales below and above it. Lower and upper dash-dotted horizontal lines denote $(2\pi)^{-1}=0.1591\ldots$ and $(2\pi)^{-1/2}=0.3989\ldots$, respectively.}
\end{figure}
\subsection{Onicescu energies}\label{SubSec_Onicescu}
Physically, Onicescu energies \cite{Onicescu1}
\begin{subequations}\label{Onicescu1D}
\begin{align}\label{Onicescu1D_R}
O_{{\bf x}_\mathtt{n}}^{(\mathtt{d})}&=\int_{\mathcal{D}_{\bf x}^{(\mathtt{d})}}\left[\rho_\mathtt{n}^{(\mathtt{d})}({\bf x})\right]^2d{\bf x}\\
\label{Onicescu1D_K}
O_{{\bf k}_\mathtt{n}}^{(\mathtt{d})}&=\int_{\mathcal{D}_{\bf k}^{(\mathtt{d})}}\left[\gamma_\mathtt{n}^{(\mathtt{d})}({\bf k})\right]^2d{\bf k}
\end{align}
\end{subequations} 
quantify a deviation of the corresponding distribution from the homogeneous one. This is easiest to show on the example of the function with compact support when the uniform distribution $\rho_{uni}^{(\mathtt{d})}$ is just a constant that is an inverse of the corresponding volume, $\rho_{uni}^{(\mathtt{d})}=1/V^{(\mathtt{d})}$. Then, the global measure of the difference between any other dependence and its uniform fellow can be described by the quantity:
$$
\int_{\mathcal{D}_{\bf x}^{(\mathtt{d})}}\left[\rho_\mathtt{n}^{(\mathtt{d})}({\bf x})-\rho_{uni}^{(\mathtt{d})}\right]^2d{\bf x}.
$$
Elementary, one obtains:
\begin{equation}\label{Onicescu2}
\int_{\mathcal{D}_{\bf x}^{(\mathtt{d})}}\left[\rho_\mathtt{n}^{(\mathtt{d})}({\bf x})-\rho_{uni}^{(\mathtt{d})}\right]^2d{\bf x}=O_{{\bf x}_\mathtt{n}}^{(\mathtt{d})}-\frac{1}{V^{(\mathtt{d})}}.
\end{equation}
Hence, the smaller the Onicescu energy is, the closer the corresponding density is to the uniform one.

As it follows from their definitions for the continuous functions, Onicescu energies are measured in units of the inverse volume of the corresponding domain; accordingly, they are not those energies that, e.g., are eigen values of the Schr\"{o}dinger equation and measured in Joules. To underline the difference, the Romanian mathematician who proposed them for the discrete probabilities
\begin{equation}\label{OnicescuDiscrete1}
O_p=\sum_{i=1}^Np_i^2,
\end{equation}
used the term 'information energy' \cite{Onicescu1}. Interchangeably and less confusing, they are also often called 'disequilibria'. Chronologically, it has to be mentioned that equation~\eref{OnicescuDiscrete1} has been used more than half a century before O. Onicescu \cite{Onicescu1} with a brief history described in ref.~\cite{Ellerman1}. There is no known rigorous proof on the limits of the product of the position and wave vector components but apparently it should not exceed that of the HO:
\begin{equation}\label{OnicescuUncertainty1}
O_{{\bf x}_\mathtt{n}}^{(\mathtt{d})}O_{{\bf k}_\mathtt{n}}^{(\mathtt{d})}\leq\frac{1}{(2\pi)^\mathtt{d}},
\end{equation}
with the equal HO contributions which for the ground orbital are:
\begin{equation}\label{OnicescuHO1}
x_\omega^\mathtt{d}O_{{\bf x}_\mathtt{0}}^{(\mathtt{d})}=x_\omega^{-\mathtt{d}}O_{{\bf k}_\mathtt{0}}^{(\mathtt{d})}=\frac{1}{(2\pi)^{\mathtt{d}/2}}.
\end{equation}

For our system, disequilibria take the form:
\begin{subequations}\label{Onicescy3}
\begin{align}\label{Onicescu3_X}
O_{x_n}(\mathfrak{a})&=\langle\rho_n\rangle_{x_n}=\int_0^\infty\rho_n^2(x)dx\\
\label{Onicescu3_K}
O_{k_n}(\mathfrak{a})&=\langle\gamma_n\rangle_{k_n}=\int_{-\infty}^\infty\gamma_n^2(k)dk
\end{align}
\end{subequations}
with the ground-level position component being:
\begin{subequations}\label{Onicescu4}
\begin{align}\label{Onicescu4_1}
O_{x_0}(\mathfrak{a})&=\frac{1}{x_{2\omega}}\frac{\Gamma\!\left(2\eta+\frac{3}{2}\right)}{2^{2\eta+1}\Gamma^2(\eta+1)}
\intertext{with its limits:}
\label{Onicescu4_0}
O_{x_0}(\mathfrak{a})&=\frac{1}{x_{2\omega}}\frac{1}{\pi^{1/2}}\left[\frac{3}{4}+\left(1-\ln2^{3/2}\right)\mathfrak{a}+\ldots\right],\,\mathfrak{a}\rightarrow0\\
\label{Onicescu4_Infinity}
O_{x_0}(\mathfrak{a})&=\frac{1}{x_{2\omega}}\frac{1}{(2\pi)^{1/2}}\left(1+\frac{1}{8\mathfrak{a}^{1/2}}+\ldots\right),\,\mathfrak{a}\rightarrow\infty.
\end{align}
\end{subequations}
Dashed line in figure~\ref{Fig_Onicescu} manifests that the increasing parameter $\mathfrak{a}$ makes the position density more homogeneous: the curve decreases from HHO value of $\frac{3}{4\pi^{1/2}}=0.4231\ldots$ to $2\omega$ HO limit of $\frac{1}{(2\pi)^{1/2}}=0.3989\ldots$. Simultaneously, wave vector component smoothly deviates from the uniformity: HHO magnitude of $0.3524\ldots$ transforms at the huge parameter into its $2\omega$ HO counterpart. Since the speed of change of $O_k$ is greater than the one of $O_x$, combined position and momentum non-homogeneity gets stronger with $\mathfrak{a}$. Qualitatively, the very similar dependencies are observed for other levels too when each component gets smaller for the higher lying states: $O_{x_{n+1}}(\mathfrak{a})<O_{x_n}(\mathfrak{a})$, $O_{k_{n+1}}(\mathfrak{a})<O_{k_n}(\mathfrak{a})$, what, of course, results in a similar behaviour of their product.

\subsection{R\'{e}nyi and Tsallis entropies}\label{SubSec_Renyi}
R\'{e}nyi \cite{Renyi1,Renyi2}
\begin{subequations}\label{Renyi1}
\begin{align}\label{Renyi1_R}
R_{{\bf x}_\mathtt{n}}^{(\mathtt{d})}(\alpha)&=\frac{1}{1-\alpha}\ln\!\left(\int_{\mathcal{D}_{\bf x}^{(\mathtt{d})}}\left[\rho_\mathtt{n}^{(\mathtt{d})}({\bf x})\right]^\alpha d{\bf x}\right)\\
\label{Renyi1_K}
R_{{\bf k}_\mathtt{n}}^{(\mathtt{d})}(\alpha)&=\frac{1}{1-\alpha}\ln\!\left(\!\int_{\mathcal{D}_{\bf k}^{(\mathtt{d})}}\left[\gamma_\mathtt{n}^{(\mathtt{d})}({\bf k})\right]^\alpha d{\bf k}\right)
\end{align}
\end{subequations}
and Tsallis \cite{Tsallis1} (or, more correctly from a historical point of view, Havrda-Charv\'{a}t-Dar\'{o}czy-Patil-Taillie-Tsallis \cite{Havrda1,Daroczy1,Patil1})
\begin{subequations}\label{Tsallis1}
\begin{align}
\label{Tsallis1_R}
T_{{\bf x}_\mathtt{n}}^{(\mathtt{d})}(\alpha)&=\frac{1}{\alpha-1}\left(1-\int_{\mathcal{D}_{\bf x}^{(\mathtt{d})}}\left[\rho_\mathtt{n}^{(\mathtt{d})}({\bf x})\right]^\alpha d{\bf x}\right)\\
\label{Tsallis1_K}
T_{{\bf k}_\mathtt{n}}^{(\mathtt{d})}(\alpha)&=\frac{1}{\alpha-1}\left(1-\int_{\mathcal{D}_{\bf k}^{(\mathtt{d})}}\left[\gamma_\mathtt{n}^{(\mathtt{d})}({\bf k})\right]^\alpha d{\bf k}\right)
\end{align}
\end{subequations}
entropies with non-negative coefficient $\alpha$ present one-parameter generalizations of the Shannon measure reducing to the latter in the limit $\alpha\rightarrow1$:
\begin{subequations}\label{RenyiLimits}
\begin{align}\label{RenyiLimits_1}
R(1)&=T(1)=S,
\intertext{as can be easily established with the help of the l'H\^{o}pital's rule. Onicescu energy is their another particular case:}
\label{RenyiLimits_2}
O&=e^{-R(2)}=1-T(2).
\intertext{Just second-order many-body R\'{e}nyi entanglement entropy of an ensemble of triphenylphosphine molecules \cite{Niknam1} and Bose-Einstein condensates of the interacting $^{87}$Rb atoms \cite{Islam1,Kaufman1} and $^{40}$Ca$^+$ ions \cite{Brydges1} was measured in recent cutting-edge experiments. In addition, R\'{e}nyi and Tsallis entropies are expressed through each other as:}
\label{RenyiTsallisRelation1_1}
T&=\frac{1}{\alpha-1}\left[1-e^{(1-\alpha)R}\right]\\
\label{RenyiTsallisRelation1_2}
R&=\frac{1}{1-\alpha}\ln(1+(1-\alpha)T).
\end{align}
\end{subequations}

Physically, introduction of the parameter $\alpha$ can be construed as a description of the deviation of the system from $\alpha=1$ equilibrium and, then, these measures quantify the intensity of the response of the structure to this deflection. At the zero value of the R\'{e}nyi/Tsallis coefficient, all random occurrences are treated on an equal footing, regardless of their actual happening, and in the opposite regime of the huge $\alpha$, the events with the highest probabilities are the only ones that contribute. As a result, both functionals $R$ and $T$ are monotonically decreasing functions of their parameter; e.g., at $\alpha=0$, the R\'{e}nyi entropy (provided it does exist) is equal to the logarithm of the volume over which it is evaluated and at $\alpha=\infty$ it is:
\begin{equation}\label{RenyiInfinite1}
R_{{\bf x},{\bf k}}(\infty)=-\ln\!\left(
\begin{array}{c}
\rho_{max}\\
\gamma_{max}
\end{array}
\right),
\end{equation}
with the subscript in the right-hand side denoting a global maximum of the corresponding distribution. Fundamental difference between the two entropies is their treatment of the combined outcome of the two random events $f$ and $g$; namely, R\'{e}nyi entropy is additive (or extensive):
\begin{equation}\label{RenyiAdditive1}
R_{fg}(\alpha)=R_f(\alpha)+R_g(\alpha),
\end{equation}
what mathematically establishes the fact that the uncertainty of the happening $f$ is not influenced by the second set, if $f$ and $g$ are the two independent occurrences. Physically, equation~\eref{RenyiAdditive1} manifests that the total information acquired from the independent distributions is the sum of its counterparts for each of them. The concept of additivity was a cornerstone requirement for the R\'{e}nyi search of the generalization of the Shannon entropy \cite{Renyi1}, which is extensive too:
\begin{equation}\label{ShannonAdditive1}
S_{fg}=S_f+S_g.
\end{equation}
Contrary, the Tsallis measure is only $\textit{pseudo}$-additive:
\begin{equation}\label{TsallisAdditive1}
T_{fg}(\alpha)=T_f(\alpha)+T_g(\alpha)+(1-\alpha)T_f(\alpha)T_g(\alpha).
\end{equation}

It is known that momentum components are defined not on the whole non-negative semi-infinite axis $0\leq\alpha<\infty$ (with the only exception of the HO) but this range is limited from below by the positive threshold $\alpha_{TH}$ \cite{Olendski2,Olendski4,Olendski6,Olendski7,Olendski8,Olendski9,Aptekarev1}. This value is determined by the system geometry, its dimensionality \cite{Olendski6,Olendski8,Olendski9,Aptekarev1} and external influences, such as, e.g., electric or magnetic fields \cite{Olendski2}. In addition, it might depend on the orbital itself \cite{Olendski2,Aptekarev1}.

Sobolev inequality of the Fourier transform \cite{Beckner2}
\begin{equation}\label{Sobolev1}
\left(\frac{\alpha}{\pi}\right)^{\mathtt{d}/(4\alpha)}\left[\int_{\mathcal{D}_{\bf x}^{(\mathtt{d})}}\rho_\mathtt{n}^\alpha({\bf x})d{\bf x}\right]^{1/(2\alpha)}\geq\left(\frac{\beta}{\pi}\right)^{\mathtt{d}/(4\beta)}\left[\int_{\mathcal{D}_{\bf k}^{(\mathtt{d})}}\gamma_\mathtt{n}^\beta({\bf k})d{\bf k}\right]^{1/(2\beta)}
\end{equation}
with the conjugation between $\alpha$ and $\beta$
\begin{equation}\label{Sobolev3}
\frac{1}{\alpha}+\frac{1}{\beta}=2
\end{equation}
straightforwardly leads to the Tsallis uncertainty relation \cite{Rajagopal1}:
\begin{equation}\label{TsallisInequality1}
\left(\frac{\alpha}{\pi}\right)^{\mathtt{d}/(4\alpha)}\!\!\left[1+(1-\alpha)T_{{\bf x}_\mathtt{n}}^{(\mathtt{d})}(\alpha)\right]^{1/(2\alpha)}\geq\left(\frac{\beta}{\pi}\right)^{\mathtt{d}/(4\beta)}\!\!\left[1+(1-\beta)T_{{\bf k}_\mathtt{n}}^{(\mathtt{d})}(\beta)\right]^{1/(2\beta)},
\end{equation}
that in the neighbourhood of $\alpha=1$ degenerates to \cite{Olendski2}:
\begin{equation}\label{TsallisInequality1_1}
L^{\frac{1-\alpha}{2\alpha}\mathtt{d}}\frac{1+\left[-2\overline{S}_{\bf x_\mathtt{n}}^{(\mathtt{d})}+\mathtt{d}(1+\ln\pi)\right](\alpha-1)/4}{\pi^{\mathtt{d}/4}}\geq L^{\frac{\beta-1}{2\beta}\mathtt{d}}\frac{1+\left[2\overline{S}_{\bf k_\mathtt{n}}^{(\mathtt{d})}-\mathtt{d}(1+\ln\pi)\right](\alpha-1)/4}{\pi^{\mathtt{d}/4}},\quad\alpha\rightarrow1.
\end{equation}
Here, $L$ is some characteristic length of the system (e.g., for our geometry, it is either $x_\omega$ or $x_{2\omega}$, as discussed above) and overlined quantities denote dimensionless Shannon entropies \cite{Olendski7}:
\begin{equation}\label{ShannonDimensionless1}
\overline{S}_{{\bf x},{\bf k}}^{(\mathtt{d})}=\mp\mathtt{d}\ln L+S_{{\bf x},{\bf k}}^{(\mathtt{d})}.
\end{equation}
Equation~\eref{TsallisInequality1_1} immediately shows that the limit $\alpha=1$ saturates Tsallis uncertainty relation what, of course, is also directly seen from equation~\eref{Sobolev1}. Besides, recalling inequality for the Shannon functionals, equation~\eqref{ShannonInequality1D}, one concludes that relation~\eref{TsallisInequality1_1} is violated at $\alpha>1$; hence, in addition to the conjugation from equation~\eref{Sobolev3}, Tsallis inequality~\eqref{TsallisInequality1} is valid in the range
\begin{equation}\label{Sobolev2}
\frac{1}{2}\leq\alpha\leq1
\end{equation}
only. Taking the logarithm of both sides of Sobolev inequality~\eref{Sobolev1} yields the R\'{e}nyi uncertainty relation \cite{Bialynicki1,Zozor1}:
\begin{equation}\label{RenyiUncertainty1}
R_{{\bf x}_\mathtt{n}}^{(\mathtt{d})}(\alpha)+R_{{\bf k}_\mathtt{n}}^{(\mathtt{d})}(\beta)\geq-\frac{\mathtt{d}}{2}\left(\frac{1}{1-\alpha}\ln\frac{\alpha}{\pi}+\frac{1}{1-\beta}\ln\frac{\beta}{\pi}\right),
\end{equation}
simultaneously removing restriction~\eref{Sobolev2}. In the limit $\alpha\rightarrow1$, it degenerates into its Shannon counterpart, equation~\eref{ShannonInequality1D}. This is another manifestation of the differences between R\'{e}nyi and Tsallis entropies. Uncertainty relations are an indispensable tool in data compression, quantum cryptography, entanglement witnessing, quantum metrology and other tasks employing correlations between the position and momentum components of the information measures \cite{Wehner1,Jizba1,Coles1,Toscano1,Hertz1,Wang1}.

As it follows from equations~\eref{Renyi1}, R\'{e}nyi entropies, similar to the Shannon ones, are measured in units of the logarithm of the distance whereas the expressions of the Tsallis measures, equations~\eref{Tsallis1}, contain a dimensional incompatibility when the term of some power of length  is subtracted from the unitless number. The reason of this ambiguity lies in the fact that definitions~\eref{Renyi1} and \eref{Tsallis1}, similar to the Shannon entropy, were directly transferred to the  continuous distributions from the discrete case where they are defined as:
\begin{subequations}\label{RenyiTsallisDiscrete1}
\begin{align}\label{RenyiTsallisDiscrete1_R}
R(\alpha)&=\frac{1}{1-\alpha}\ln\!\!\left(\sum_{n=1}^Np_n^\alpha\right)\\
\label{RenyiTsallisDiscrete1_T}
T(\alpha)&=\frac{1}{\alpha-1}\left(1-\sum_{n=1}^Np_n^\alpha\right).
\end{align}
\end{subequations}
Despite this incompatibility, Tsallis inequality~\eqref{TsallisInequality1} is dimensionally correct what, in particular, is exemplified by equation~\eref{TsallisInequality1_1} with the use of conjugation~\eref{Sobolev3}:
\begin{equation}\label{Beta1}
\beta=\frac{\alpha}{2\alpha-1}.
\end{equation}

Let us start our discussion of the 1D PHO from the dependence of the threshold coefficient $\alpha_{TH}$ on the parameter $\mathfrak{a}$. For simplicity, consider the ground-state momentum R\'{e}nyi entropy:
\begin{equation}\label{RenyiK0}
R_{k_0}(\alpha)=\frac{1}{1-\alpha}\ln\!\left(\int_{-\infty}^\infty\gamma_0^\alpha(k)dk\right),
\end{equation}
where, upon using the density $\gamma_0(k)$, according to equations~\eref{Density1_k} and \eref{MomentumFunction0}, one has to evaluate the logarithm of the following integral:
\begin{equation}\label{Integral2}
\int_0^\infty\!\left(\!e^{-y^2}\!\left[\Gamma^2\!\left(\frac{\eta}{2}+\frac{3}{4}\right)\!M^2\!\left(-\frac{\eta}{4}-\frac{1}{4};\frac{1}{2};\frac{y^2}{2}\right)+2\Gamma^2\!\left(\frac{\eta}{2}+\frac{5}{4}\right)\!y^2M^2\!\left(-\frac{\eta}{2}+\frac{1}{4};\frac{3}{2};\frac{y^2}{2}\right)\right]\right)^\alpha\!\!dy.
\end{equation}
To understand under which conditions this improper integral does converge, one employs asymptotic expansion of the Kummer confluent hypergeometric functions \cite{Abramowitz1} to find out that at infinity both terms in integral~\eref{Integral2} have the same $y$ dependence:
\begin{subequations}
\begin{align}
e^{-y^2}M^2\left(-\frac{\eta}{4}-\frac{1}{4};\frac{1}{2};\frac{y^2}{2}\right)\rightarrow\frac{1}{y^{2\eta+3}},\quad y\rightarrow\infty\\
e^{-y^2}y^2M^2\left(-\frac{\eta}{2}+\frac{1}{4};\frac{3}{2};\frac{y^2}{2}\right)\rightarrow\frac{1}{y^{2\eta+3}},\quad y\rightarrow\infty.
\end{align}
\end{subequations}
Accordingly, the integral will converge at $(2\eta+3)\alpha>1$ what immediately leads to the following expression for the threshold:
\begin{subequations}\label{Threshold1}
\begin{align}\label{Threshold1_1}
\alpha_{TH}&=\frac{1}{3+\sqrt{1+4\mathfrak{a}}}
\intertext{with asymptotes}
\label{Threshold1_0}
\alpha_{TH}&=\frac{1}{4}-2\mathfrak{a}+\frac{3}{16}\,\mathfrak{a}^2+\ldots,\quad\mathfrak{a}\rightarrow0\\
\label{Threshold1_Infinity}
\alpha_{TH}&=\frac{1}{2\mathfrak{a}^{1/2}}-\frac{3}{4\mathfrak{a}}+\ldots,\quad\mathfrak{a}\rightarrow\infty.
\end{align}
\end{subequations}
It is easy to show that this fundamental relation is a level-independent one; indeed, at any $n$, for example, the real part  $\Phi_n^{(r)}(\mathfrak{a};k)$ of the waveform will contain the sum of the functions $M\!\left(-\frac{\eta}{4}-\frac{1}{4}-j;\frac{1}{2};\frac{y^2}{2}\right)$ with $j=0,1,\ldots,n$, what will result in the $y^{-(4j+2\eta+3)}$ dependence at infinity. This, in turn, means that the term with $j=0$ will determine the overall convergence. Very similar reasoning applies to the imaginary parts $\Phi_n^{(i)}(\mathfrak{a};k)$ too. Equations~\eref{Threshold1} promulgate that the HHO lower edge is equal to one quarter and, as the PHO coefficient $\mathfrak{a}$ grows, it decreases to zero when the potential takes the form of $2\omega$ HO. As it is known \cite{Olendski4}, the HO is the only structure for which the momentum functional, similar to it position counterpart, exists at any non-negative $\alpha$. Obviously, the same conclusions apply to the Tsallis measures too.

\begin{figure}
\centering
\includegraphics[width=\columnwidth]{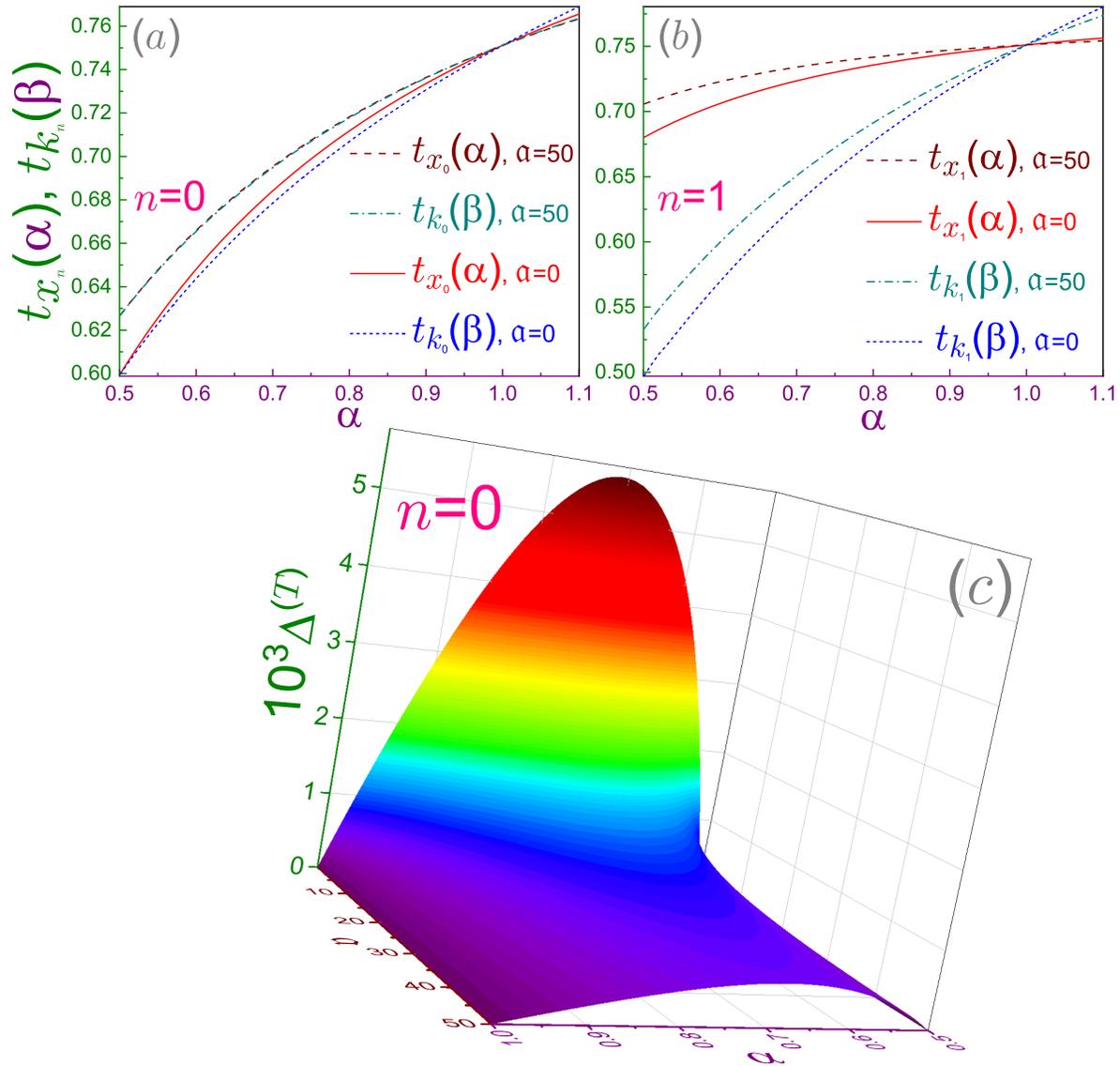}
\caption{\label{Fig_Tsallis}
(a) Ground-state, $n=0$, dimensionless Tsallis position $t_{x_n}(\mathfrak{a};\alpha)$ functions for HHO and PHO with $\mathfrak{a}=50$ (solid and dashed line, respectively) together with their wave vector  counterparts  $t_{k_n}(\mathfrak{a};\beta)$ (dotted and dash-dotted curve, respectively). Due to small difference between $t_{x_0}$ and $t_{k_0}$ at $\mathfrak{a}=50$, the corresponding dependencies are almost unresolved in the scale of the figure. (b) The same as in window (a) but for the first excited state, $n=1$. Note different vertical ranges and scales in these two subplots. (c) Quantity $\Delta^{(T)}$, equation~\eref{TsallisD1}, magnified one thousand times, in terms of PHO factor $\mathfrak{a}$ and Tsallis parameter $\alpha$.}
\end{figure}

Among many other properties of the 1D PHO entropies, we will discuss  below the corresponding uncertainty relations. General Tsallis inequality~\eqref{TsallisInequality1} in our case reads:
\begin{equation}\label{TsallisInequality2}
x_\omega^\frac{1-\alpha}{2\alpha}t_{x_n}(\alpha)\geq x_\omega^\frac{\beta-1}{2\beta}t_{k_n}(\beta),
\end{equation}
where explicit expressions of the dimensionless position $t_{x_n}(\alpha)$ and momentum $t_{k_n}(\beta)$ Tsallis components due to their unwieldiness are not written here. As already discussed above, this equation is dimensionally correct. Figure~\ref{Fig_Tsallis}(a) shows $t_{x_0}(\alpha)$ and $t_{k_0}(\beta)$ for HHO and $\mathfrak{a}=50$. It confirms our earlier conclusion that the Tsallis inequality holds true only inside the range from equation~\eref{Sobolev2} and it turns into the trivial identity at $\alpha=1$. Remarkably, the ground orbital saturates it at the opposite edge of one half too. This phenomenon was discovered before \cite{Olendski2,Olendski4,Olendski6,Olendski7,Olendski8,Olendski9} and its PHO explanation is very similar to other structures \cite{Olendski2,Olendski6,Olendski7,Olendski8,Olendski9}; namely, at $\alpha=1/2$, left-hand side of relation~\eref{TsallisInequality1} or, equivalently, \eref{Sobolev1}, turns to
\begin{equation}\label{Integral3}
\frac{1}{(2\pi)^{1/2}}\int_0^\infty|\Psi_n(\mathfrak{a};x)|dx,
\end{equation}
whereas in the right-hand one its coefficient $\beta$ tending to infinity picks up from the integral the maximum of the momentum waveform $\left|\Phi_n(\mathfrak{a};k)\right|_{max}$. Next, in equation~\eref{Integral3} the absolute value of the function can be replaced by the function itself for the ground level only, $n=0$, since $\Psi_0(\mathfrak{a};x)$ is a sole position dependence that does not have nodes. Then, from the Fourier transform~\eref{Fourier1} one gets:
\begin{equation}\label{Integral4}
\frac{1}{(2\pi)^{1/2}}\int_0^\infty\Psi_0(\mathfrak{a};x)dx=\Phi_0(\mathfrak{a};0).
\end{equation}
Zero wave vector is an extremum of the corresponding distribution for any orbital and, as figure~\ref{Fig_MomentumWaveform0} demonstrates, it is a global maximum of $\Phi_0(\mathfrak{a};k)$. This concludes a proof of the statement that at $\alpha=1/2$ the ground state converts Tsallis relation into the identity. To develop it further, panel (b) plots the same dependencies for the first excited orbital. It shows that the corresponding curves intersect at $\alpha=1$ only whereas at any smaller parameter, including the left edge of the interval~\eref{Sobolev2}, position part is strictly greater than the momentum component. In addition, it is important to point out that at the increasing PHO factor, the difference between $t_{x_0}(\alpha)$ and $t_{k_0}(\beta)$ shrinks; for example, at $\mathfrak{a}=50$, the  designated lines are practically unresolved in the scale of panel (a). To elaborate on this, window (c) exhibits the difference
\begin{equation}\label{TsallisD1}
\Delta^{(T)}=t_{x_0}(\alpha)-t_{k_0}(\beta),
\end{equation}
which monotonically dwindles unless at the infinitely large $\mathfrak{a}$ it turns to zero on  the whole semi-axis $\alpha\geq1/2$: for the HO ground-state, irrespectively of the Tsallis coefficient, dimensionless part of each side is equal to $\pi^{-1/4}$ with the constraint from requirement~\eref{Sobolev2} being waived \cite{Olendski4}. This is another manifestation of a special cachet of the Gaussian distribution \cite{DePalma1}.

Switching to the R\'{e}nyi entropy again, let us note first that its ground-state position component is calculated analytically:
\begin{subequations}\label{RenyiPosition0}
\begin{align}\label{RenyiPosition0_1}
R_{x_0}(\mathfrak{a};\alpha)-\ln x_\omega&=-\ln2+\frac{1}{1-\alpha}\ln\frac{\Gamma\left(\alpha\!\left(\eta+\frac{1}{2}\right)+\frac{1}{2}\right)}{\Gamma^\alpha(\eta+1)\alpha^{\alpha\!\left(\eta+\frac{1}{2}\right)+\frac{1}{2}}}.
\intertext{For the vanishing coefficient, it reads:}
\label{RenyiPosition0_0}
R_{x_0}(\mathfrak{a};\alpha)-\ln x_\omega&=-\ln2+\frac{1}{2}\ln\pi-\frac{1}{2}\ln\alpha+\ldots,\quad\alpha\rightarrow0,
\intertext{and for the infinite R\'{e}nyi factor it is:}
\label{RenyiPosition0_Infinity}
R_{x_0}(\mathfrak{a};\infty)-\ln x_\omega&=-\ln2+\eta+\frac{1}{2}-\ln\!\left(\frac{1}{\Gamma(\eta+1)}\left(\eta+\frac{1}{2}\right)^{\eta+\frac{1}{2}}\right).
\intertext{Divergence at the zero $\alpha$, which in this regime does not depend on the PHO parameter, is explained by the semi-infinite range of integration. Relation~\eref{RenyiPosition0_Infinity} was derived from equation~\eref{RenyiInfinite1} by zeroing first a derivative of the corresponding expression of $\rho_0(x)$  and finding in this way the location $\left(\eta+\frac{1}{2}\right)^{1/2}x_\omega$ of the maximum (note that at $\mathfrak{a}\rightarrow\infty$ it approaches minimum of the potential, equation~\eref{PositionZero1}, as expected) and then substituting it back into $\rho_0(x)$. Also, in the limit $\alpha\rightarrow1$ R\'{e}nyi entropy degenerates into its Shannon counterpart~\eref{PositionShannon0_1}:}
\label{RenyiPosition0_Shannon}
R_{x_0}(\mathfrak{a};\alpha)=S_{x_0}(\mathfrak{a})&+\left[\frac{1}{2}+\frac{1}{\eta}+\frac{1}{8\eta^2}-\frac{1}{2}\left(\eta+\frac{1}{2}\right)^2\psi^{(1)}(\eta)\right](\alpha-1)+\ldots,\quad\alpha\rightarrow1,
\intertext{$\psi^{(j)}(z)=d^j\psi(z)/dz^j$, $j=1,2,\ldots$, is a polygamma function \cite{Abramowitz1}, and at $\alpha=2$ its exponent turns, according to relation~\eref{RenyiLimits_2}, into the Onicescu energy, equation~\eref{Onicescu4_1}. Since the R\'{e}nyi functional is a monotonically decreasing function of its parameter, expression in the square brackets in equation~\eref{RenyiPosition0_Shannon} is negative at any $\mathfrak{a}$ what can be easily checked by direct calculation. It is also possible to derive the limiting cases of the potential; namely, near-HHO geometry yields}
R_{x_0}(\mathfrak{a};\alpha)-\ln x_\omega&=-\ln2+\frac{1}{1-\alpha}\ln\frac{\Gamma\left(\alpha+\frac{1}{2}\right)}{\left(\frac{\pi^{1/2}}{2}\right)^\alpha\alpha^{\alpha+\frac{1}{2}}}\nonumber\\
\label{RenyiPosition0_a0}
&+\frac{\alpha}{1-\alpha}\left[\psi\left(\alpha+\frac{1}{2}\right)-2+\gamma+\ln\frac{4}{\alpha}\right]\mathfrak{a}+\ldots,\quad\mathfrak{a}\rightarrow0,
\intertext{and the functional approaches its $2\omega$ HO counterpart $R_{x_0}^{(2\omega)}(\alpha)=\ln x_{2\omega}+\frac{1}{2}\ln\pi-\frac{1}{2}\frac{\ln\alpha}{1-\alpha}$ \cite{Olendski4} by obeying the rule:}
\label{RenyiPosition0_aInfinity}
R_{x_0}(\mathfrak{a};\alpha)&=R_{x_0}^{(2\omega)}(\alpha)-\frac{\alpha+1}{24\alpha}\left(\frac{1}{\mathfrak{a}^{1/2}}-\frac{1}{2\mathfrak{a}}+\ldots\right),\quad\mathfrak{a}\rightarrow\infty.
\end{align}
\end{subequations}
In deriving the last dependence, a crucial role was played by Stirling's formula \cite{Abramowitz1}:
$$\ln\Gamma(z)\sim\left(z-\frac{1}{2}\right)\ln z-z+\frac{1}{2}\ln2\pi+\frac{1}{12z}+\ldots,\quad z\rightarrow\infty.$$
It is instructive to point out that the unit value of the R\'{e}nyi parameter, $\alpha=1$, transforms equation~\eref{RenyiPosition0_aInfinity} into its Shannon simplification, equation~\eref{PositionShannon0_Infinity}. The same is true about equations~\eref{RenyiPosition0_a0} and \eref{PositionShannon0_0}, respectively, where however, contrary to the previous link, the l'H\^{o}pital's rule is applied.

\begin{figure}
\centering
\includegraphics[width=0.8\columnwidth]{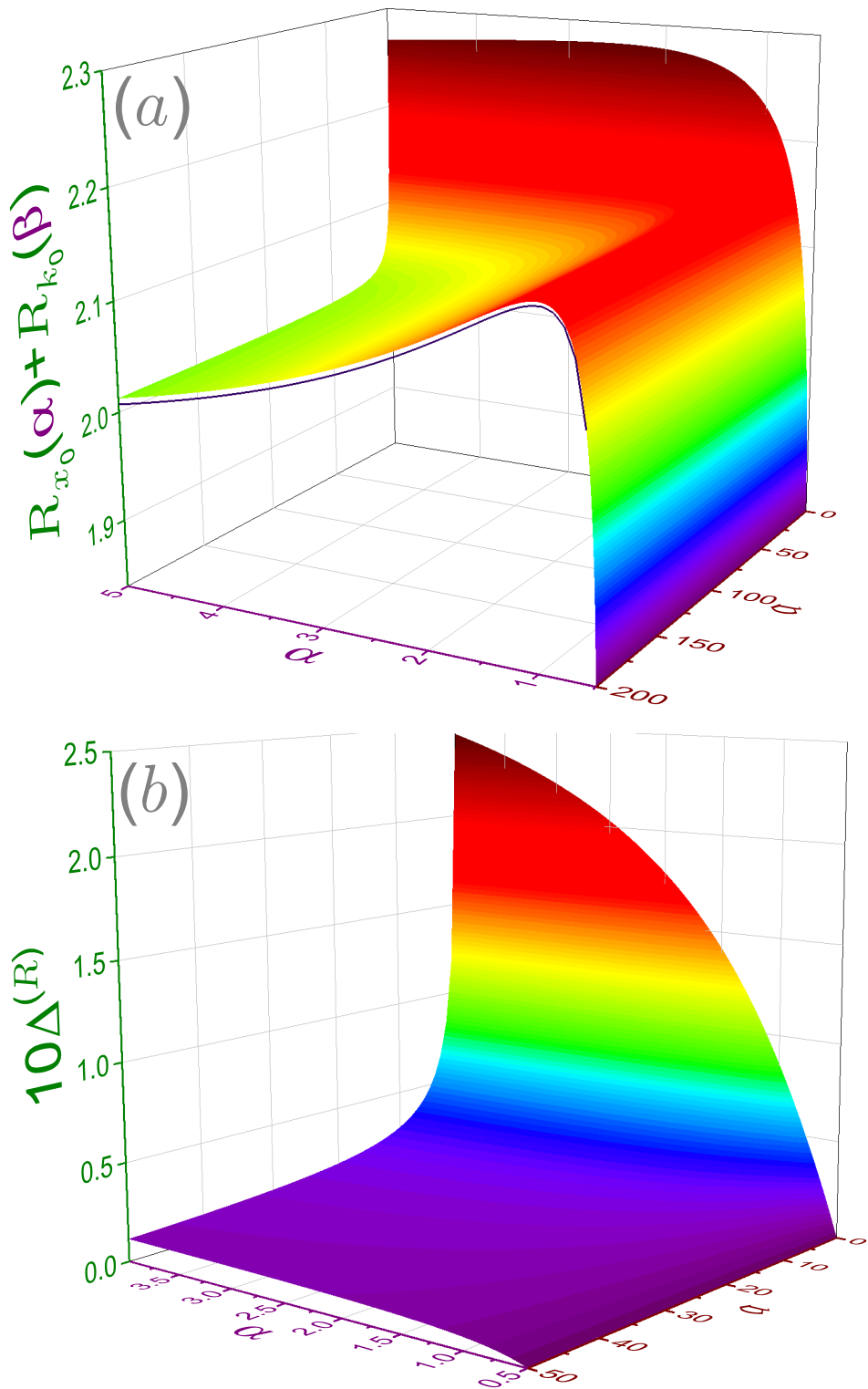}
\caption{\label{Fig_Renyi}
(a) Ground-state left-hand side $R_{x_0}(\alpha)+R_{k_0}(\beta)$ of R\'{e}nyi uncertainty relation~\eref{RenyiUncertainty1} in terms of PHO factor $\mathfrak{a}$ and R\'{e}nyi parameter $\alpha$. Dark curve at the front panel depicts right-hand side of the same equation. (b) Quantity $\Delta^{(R)}$, equation~\eref{RenyiD1}, magnified ten times, in terms of $\mathfrak{a}$ and $\alpha$.}
\end{figure}

Figure~\ref{Fig_Renyi}(a) exhibits a dependence of the left-hand side of the ground-state R\'{e}nyi uncertainty relation~\eref{RenyiUncertainty1} in terms of the PHO $\mathfrak{a}$ and R\'{e}nyi $\alpha$ parameters. At any $\mathfrak{a}$, this inequality turns into the identity at $\alpha=1/2$ when its either part is equal to $\ln2\pi=1.8378\ldots$. Explanation of this is the same as for the Tsallis entropy presented above and is not discussed here. $R_{x_0}(\alpha)+R_{k_0}(\beta)$ as a function of the coefficient $\alpha$ possesses HHO broad maximum of $2.280\ldots$ located at approximately $2.55$. As the potential shape changes, this maximum shrinks narrower simultaneously decreasing and shifting to the smaller $\alpha$ until at $\mathfrak{a}\rightarrow\infty$ the corresponding dependence turns to the HO one when the maximum of $1+\ln\pi$ is located at the Shannon case \cite{Olendski4}. For comparison, the HO dependence is shown by the black line at the front panel. To underline the convergence of the uncertainty relation~\eref{RenyiUncertainty1} into the identity with the change of the shape, panel (b) depicts the difference $\Delta^{(R)}$ between its left- and right-hand sides:
\begin{equation}\label{RenyiD1}
\Delta^{(R)}=R_{x_0}(\alpha)+R_{k_0}(\beta)+\frac{\mathtt{1}}{2}\left(\frac{1}{1-\alpha}\ln\frac{\alpha}{\pi}+\frac{1}{1-\beta}\ln\frac{\beta}{\pi}\right).
\end{equation}
Of course, it is equal to zero at $\alpha=1/2$ and any potential profile. It is seen also that at $\alpha>1/2$ the difference get smaller with the increase of $\mathfrak{a}$ until at $\mathfrak{a}=\infty$ it identically turns to zero: for the HO, the ground-state uncertainty relation has its both sides equal to each other at any R\'{e}nyi parameter $\alpha$ \cite{Olendski4}.

\section{Concluding remarks}\label{Sec_Conclusions}
Consideration of the models where a variation of one parameter leads to the change of the physical characteristics of the system enriches our knowledge from scientific and technological perspectives. One of such examples is introduced and scrutinized in the present research where a modification of the 1D potential profile is controlled by the dimensionless coefficient $\mathfrak{a}$ whose growth transforms asymmetric HHO into the regular HO with double frequency. Such evolution was investigated here with the emphasis on the analysis of quantum information measures. Some quantities, e.g., standard deviations and associated Heisenberg uncertainty relations or position components  - either all, as for the Fisher information, or, at least, ground-state ones, as for the Shannon and R\'{e}nyi entropies - allow analytic mathematical expressions whose asymptotes shed a bright light on the way they approach their HO counterparts. For the vast majority of the momentum components, numerical calculations are the only possible source of getting information but even here it has to be mentioned that a method was proposed for obtaining the analytic form of the corresponding waveforms what allowed, in particular, to derive a simple formula for the lower boundary that defines the range of the existence of the R\'{e}nyi or Tsallis functionals. Exposition pays a large attention to the physical interpretation of the obtained results.

With contemporary nanotechnology advances, it is possible to grow quantum structures of any desired geometry. Accordingly, it should be no difficulty to build up 1D systems whose profile is described by equation~\eref{Potential1D_1} and to investigate their properties at different $\mathfrak{a}$. Alternatively, at the fixed parameter one can vary the position of the minimum with respect to the left border by introducing an electric field $\pmb{\mathcal{E}}$ directed along the $x$ axis what experimentally is achieved by, e.g., applying the appropriate gate voltage. Mathematically, an insertion of the item $-|e|\mathcal{E}x$ into the right-hand side of equation~\eref{Potential1D_1} modifies solutions of the Schr\"{o}dinger equation: now they are expressed with the help of Heun functions \cite{Ronveaux1,Hortacsu1} - but physically its influence leads to the shift of the potential extremum away from the wall what results in the transformation of the structure into the 1D HO.

Finally, let us point out again that proposed here model of the modified PHO with its shape being controlled by the parameter $\mathfrak{a}$ has been widely used, as mentioned in the Introduction, in the analysis of the 2D quantum rings \cite{Bogachek1,Tan1,Tan2,Tan3,Fukuyama1,Bulaev1,Simonin1,Olendski5,Gumber1,Olendski1,Olendski2}. It is natural to expand such approach of investigating quantum-information concepts to any other dimension $\mathtt{d}\geq3$. For the 3D structure, the measures $S$, $I$, $O$, $R(\alpha)$ and $T(\alpha)$  have been calculated at $\mathfrak{a}=1$ only \cite{Yahya1} and for any higher dimensionality just energy spectrum and position wave functions have been analyzed \cite{Wang2,Oyewumi2,Oyewumi3,Das1}.
\ack
Research was supported by Competitive Research Project No. 2002143087 from the Research Funding Department, Vice Chancellor for Research and Graduate Studies, University of Sharjah.

\end{document}